\documentclass[aps,prd,reprint,superscriptaddress,nofootinbib,longbibliography]{revtex4-1}
\usepackage{amssymb,amsmath,bm,natbib}
\usepackage[dvipsnames]{xcolor}
\usepackage{microtype}
\usepackage{graphicx}
\usepackage{xspace}
\usepackage{bm}
\usepackage{bbm}
\usepackage{tikz}
\usepackage{dsfont}
\usetikzlibrary{decorations.pathmorphing}
\usetikzlibrary{decorations.markings}
\usetikzlibrary{arrows}
\tikzset{snake it/.style={decorate, decoration=snake}}
\usetikzlibrary{matrix,calc}
\usepackage{upgreek}
\usepackage[colorlinks = true,
            linkcolor = Maroon,
            urlcolor  = Maroon,
            citecolor = Maroon]{hyperref}

\usepackage{color}
\usepackage{chemformula} 
\begin{document}
\title{The $\beta$-decay spectrum of Tritiated graphene: combining nuclear quantum mechanics with Density Functional Theory} 
\author{Andrea~Casale}
\email{andrea.casale@columbia.edu}
\affiliation{Department of Physics, Columbia University, New York, 538
West 120th Street, NY 10027, USA}
\author{Angelo~Esposito}
\email{angelo.esposito@uniroma1.it}
\affiliation{Dipartimento di Fisica, Sapienza Universit\`a di Roma, Piazzale Aldo Moro 2, I-00185 Rome, Italy}
\affiliation{INFN Sezione di Roma, Piazzale Aldo Moro 2, I-00185 Rome, Italy}
\author{Guido~Menichetti}
\email{guido.menichetti@df.unipi.it}
\affiliation{Dipartimento di Fisica dell’Universit\`a di Pisa, Largo Bruno Pontecorvo 3, I-56127 Pisa, Italy}
\affiliation{Fondazione Istituto Italiano di Tecnologia, Center for Nanotechnology Innovation@NEST, Piazza San Silvestro 12, I-56127 Pisa, Italy}
\author{Valentina~Tozzini}
\email{valentina.tozzini@nano.cnr.it}
\affiliation{Istituto Nanoscienze - CNR, Lab NEST-SNS, Piazza San Silvestro 12, I-56127 Pisa, Italy}
\affiliation{INFN Sezione di Pisa, Largo Bruno Pontecorvo 3, I-56127 Pisa, Italy}
\date{\today}
\begin{abstract}
\noindent We present the results of a multi-methodological study aimed at investigating the interaction between graphene and Tritium during its $\beta$-decay to Helium, under different levels of loading and geometrical configurations. We combine Density Functional Theory (DFT), to evaluate the interaction potentials, with calculations of the decay rate, in order to study the consequences that the presence of the substrate has on the $\beta$-decay spectrum of Tritium. We determine the shape of the event rate, accounting for the effects of (part of) the corresponding condensed matter degrees of freedom. In the context of future neutrino experiments, our results provide important information aimed at the optimization of hosting material. Furthermore, our work outlines a novel theoretical and computational scheme to account for the different time scales involved in the process, and to address a question at the boundary between high and low energy physics. This requires non-conventional declinations of DFT combined with full quantum treatments of the nuclear configuration involved in the decay process. Our results pave the way for future studies, aimed at determining the physics reach of upcoming neutrino mass experiments.
\end{abstract}
\maketitle
\section{Introduction}
\noindent
Of all the known fundamental particles, neutrinos are arguably the most elusive ones, mostly because, aside from gravity, they couple to the rest of ordinary matter only via weak interactions, which, at low energies, are extremely feeble. Moreover, while within the renormalizable part of the Standard Model they are exactly massless, the experimental observation of the oscillations between their three flavors (i.e., electronic, muonic and tauonic), indeed imply that they do carry mass~\cite{Super-Kamiokande:1998kpq,SNO:2001kpb,SNO:2002tuh,deSalas:2020pgw,Esteban:2020cvm,Capozzi:2021fjo}. These masses, however, are still unknown, both in their absolute value (i.e., the mass of the lightest neutrino), and in their ordering (i.e., whether the lightest neutrino is mostly within the electronic or tauonic family). 

A possible way to measure these properties is by looking at the spectrum of the electrons emitted by $\beta$-decay processes of unstable nuclei. Among them, a notable instance is the decay of Tritium (hereafter also noted as T),
\begin{align} \label{eq:decay}
    {\rm T} \to {^3}{\rm He}+ e + \bar \nu_e \,,
\end{align}
which produces an electron, an anti-neutrino and a Helium nucleus. The shape of the spectrum for values of the electron's energy close to the maximum allowed one --- the so-called {\it end-point} --- is the most sensitive to the neutrino masses, which can then in principle be inferred from its precise measurement. The Tritium lifetime of about 12 years, and its relatively low energy end-point of 18.6~keV, makes it particularly amenable to this aim, motivating a series of Tritium $\beta$-decay experiments~\cite[e.g.,][]{Robertson:1991vn,Kawakami:1991th,Holzschuh:1992np,Stoeffl:1995wm}. The current best laboratory bound on the neutrino (effective) mass has been set by the KATRIN experiment, which obtained $m_\nu < 0.45$~eV, using gaseous molecular Tritium~\cite{KATRIN:2019yun,KATRIN:2021uub,Aker:2024drp}.

Other proposed projects like PTOLEMY~\cite{Betti:2018bjv,PTOLEMY:2019hkd,Apponi:2021hdu,PTOLEMY:2022ldz,PTOLEMY:2025unk} or Project8~\cite{Project8:2022wqh,Project8:2022hun} aim at improving this bound by employing Tritium in different forms. In particular, PTOLEMY plans to distribute it on a solid substrate, with the advantage of achieving extremely high concentrations, thus maximizing the number of observed events. Albeit with additional difficulties, a similar strategy could also allow the detection of the elusive cosmic neutrino background, via the neutrino capture process~\cite{Weinberg:1962zza,Cocco:2007za,PTOLEMY:2019hkd}, i.e.,
\begin{align}
    \nu_e + {\rm T} \to {^3}{\rm He} + e \,.
\end{align}
The substrate of choice has been identified as graphene, whose capability of covalently binding Hydrogen~\cite{cami15} in almost 1:1 stoichiometry with Carbon was recently demonstrated for nanoporous samples, displaying up to $90\%$ loading~\cite{betti22}. Moreover, its low dimensional nature allows to minimize uncontrolled energy losses for the outgoing electron~\cite{apponi2024transmission}. Thanks to this, the project has the potential of reaching an unprecedented sensitivity to the effective neutrino mass (see, e.g.,~\cite{PTOLEMY:2019hkd} for an estimate). 

\begin{figure*}
\includegraphics[width=\linewidth]{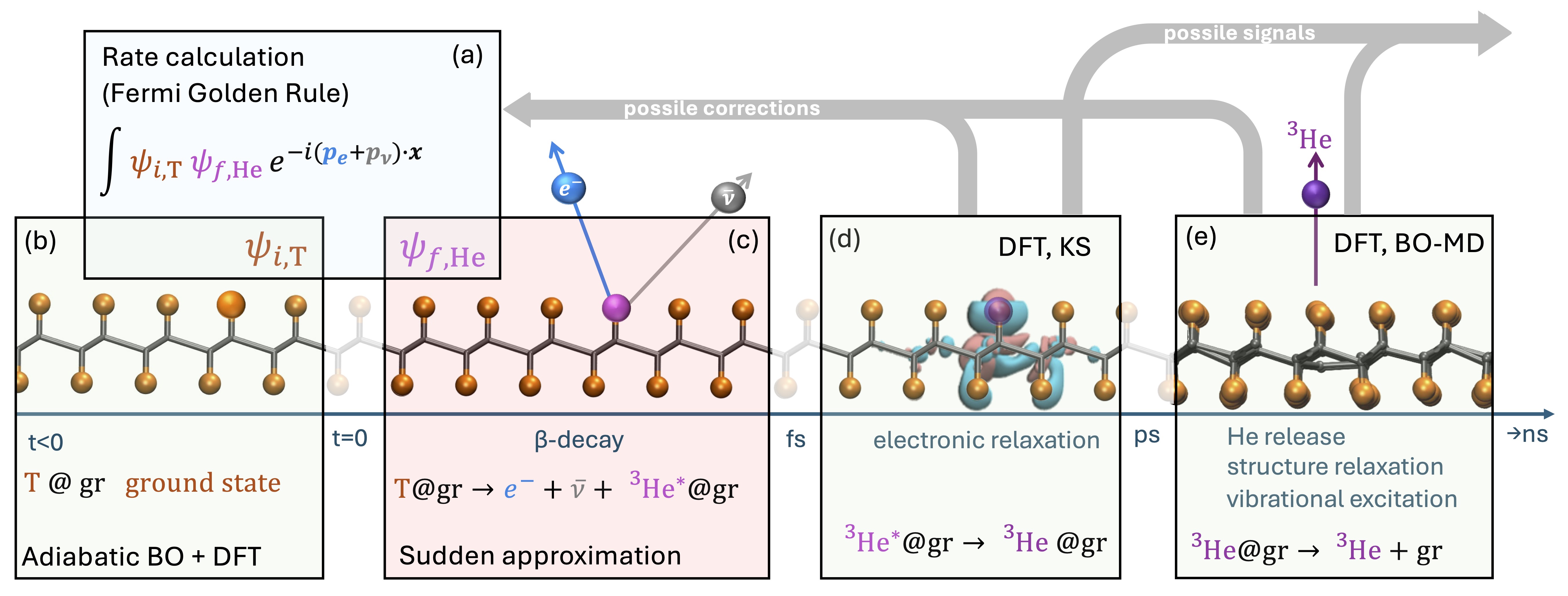}
    \caption{Scheme of the theoretical framework and calculations performed in this work. {\bf Panel (a):} Calculation of the matrix element between initial and final states to evaluate the decay rate spectrum. {\bf Panel (b):} DFT calculation of the Tritium potential on graphene in the initial ground state. {\bf Panel (c):} Sudden approximation for the calculation of the potential felt by Helium just after the decay. {\bf Panel (d):} DFT calculation of the electronic relaxation via the Kohn-Sham (KS) scheme. {\bf Panel (e):} Born-Oppenheimer dynamics (BO-MD) of the system after the electronic relaxation. Shorthand notations: T is tritium $^3$H, the asterisk $*$ indicates an electronically excited state and the suffix @gr the bond to graphene.}\label{fig:scheme}
\end{figure*}

Hydrogen (and Tritium) chemisorption alters the structural, chemical and electronic properties of the material, e.g., turning graphene from a conductor to a semiconductor~\cite{betti22}, or even an insulator in the fully loaded case, called ``graphane''.
In order to determine the possible experimental reach, it is however crucial to understand the effects that solid state phenomena have on the $\beta$-process. These can be essentially classified in two: (i) by localizing Tritium in the vicinity of a Carbon site, its wave function will not be a simple momentum eigenstate anymore, and (ii) the degrees of freedom of the substrate will now actively participate in the reaction. This affects the electron spectrum, making it qualitatively different from the one predicted for the process in vacuum. Together with the high energy resolution expected for PTOLEMY, this introduces conceptual and practical challenges for both the neutrino mass measurement, and for the detection of cosmic neutrinos~\cite{Cheipesh:2021fmg,Nussinov:2021zrj,Tan:2022eke,PTOLEMY:2022ldz,Cavoto:2022xwo}.\footnote{Conversely, it could provide an advantage in looking for keV-scale sterile neutrinos~\cite{chung25}.}

From the theoretical point of view, understanding the effects of the substrate on the energy spectra involves combining, in unconventional ways, very different theoretical approaches and numerical techniques, as those used in two domains which are typically considered as separated: the high energy domain of elementary particles and the low energy domain of solid state physics~\cite{meni25}. This can be understood considering the temporal sequence of events depicted in Fig.~\ref{fig:scheme}. The evaluation of the spectrum involves the calculation of the matrix element between the initial state and the state just after the decay (panel (a)). The state before the decay can be considered as an electronic ground state of Tritium (panel (b)), a configuration that has been well characterized within the last decades~\cite{delfino_24,bellucci20,cami15}. In particular, the theoretical studies usually rely on Density Functional Theory (DFT), capable of accurately determining the ground state properties of graphene in interaction with other elements~\cite{rossi15,cavallucci18}, and, therefore, the potential felt by Tritium on graphene. In light of this, here we use the standard DFT scheme~\cite{kohn1965self,giannozz_01}, based on the adiabatic approximation for the electrons, to describe the initial state of the system. 
Similarly, standard DFT can be used to evaluate the state of the system long after the decay, at time scales that are longer than those associated to the relaxation of the electronic structure (panel~(d)), including the long time scale dynamics of the whole system (e.g., the vibrational excitation of the substrate, panel~(e)). 

Nonetheless, the extreme non-adiabaticity of the ${\rm T} \to {\rm He}$ transition calls for non-conventional schemes to treat the state right after the decay (panel (c)). In particular, we combine the DFT scheme with the so-called ``sudden approximation'' \cite[e.g.,][]{hedin2002sudden}, treating the electronic structure as frozen,  we introduce its weaker variant, the ``semi-sudden approximation'', and develop novel dedicated tools to implement them. Using these schemes, we calculate the interaction potentials at different stages of the reaction, and use them to compute the spectrum of the outgoing electron, accounting for the possible excitations in the final Helium. Our innovative multi-methodological approach combines advanced applications of DFT together with a full quantum treatment of the nuclear states, used for the determination of the decay spectra. This poses the basis for the in-depth study of $\beta$-decay in the presence of a solid substrate, which is both conceptually new and phenomenologically relevant for future neutrino experiments. 

This manuscript is organized as follows. In Section~\ref{sec:betaintro} we describe the procedure to calculate the $\beta$-decay rate in presence of a substrate. Section~\ref{sec:initial} reports the calculations of the interaction potentials between the initial state Tritium and the substrate, exploiting a standard DFT scheme. Section~\ref{sec:final}, instead, describes the interaction of Helium with the substrate right after the decay, exploiting the sudden approximation, and also reporting a detailed explanation of the non-conventional DFT calculations needed in this context.  Finally, in Section~\ref{sec:finalrate} we discuss the actual calculation of the decay rate including the effects of the binding potentials of both Tritium and Helium,  followed by Section~\ref{sec:fate}, describing the dynamical evolution on the longer time scales after the decay. We conclude with a summary, and a discussion of the limitations and perspectives of our approach. 

\noindent{\it Conventions:} Unless otherwise specified, we work in natural units,  $\hbar = c = 1$, as well as Gaussian units, $4\pi \epsilon_0 = 1$.
\section{General approach to the $\beta$-decay rate} \label{sec:betaintro}
\noindent
When a Tritium nucleus is bound to graphene, its physical state is described by a wave function, say $\psi_{i,\rm T}(\bm x)$, belonging to the discrete spectrum of its binding potential, with $\bm x$ being its position. After the decay has happened, the outgoing electron and anti-neutrino will belong to the continuous spectrum, and their wave functions can be approximated by simple plane waves,\footnote{Since the electron is charged, there might actually be long distance Coulomb effects which distort its wave function. The inclusion of these requires the knowledge of the electric field generated by the Tritiated graphene layer at long distances, and it is beyond the scope of the present work.} $\psi_e(\bm x) = e^{i \bm p_e \cdot \bm x}/\sqrt{V}$ and $\psi_\nu(\bm x) = e^{i \bm p_\nu \cdot \bm x}/\sqrt{V}$, where we assume to be working at some finite volume, $V$. The state of the final Helium nucleus, instead, will be described by a wave function, $\psi_{f,{\rm He}}(\bm x)$, which can belong to either the discrete or the continuous spectrum of the final state Hamiltonian.

At long enough distances, the weak interaction Hamiltonian which induces the $\beta$-decay is local, and its expression in position basis is given by,
\begin{align}
    \langle \bm x_{\rm He}, \bm x_e, \bm x_\nu | H_{\rm w} | \bm x_{\rm T} \rangle ={}& g \, \delta^{(3)}(\bm x_{\rm He} - \bm x_{\rm T}) \, \delta^{(3)}(\bm x_{e} - \bm x_{\rm T}) \notag \\
    &\! \! \! \times \delta^{(3)}(\bm x_{\rm \nu} - \bm x_{\rm T}) \,,
\end{align}
with obvious definition of the position vectors. The coupling $g$ is related to the microscopic theory of weak interactions, as reported, for example, in~\cite{Simkovic:2007yi,PTOLEMY:2022ldz}.

In light of this, the $\beta$-decay event rate for a single Tritium can be obtained from Fermi's golden rule as,
\begin{align} \label{eq:dgamma}
    d\Gamma = 2\pi \sum_f |\mathcal{M}_f|^2 \delta(E_i \!-\! E_f \!-\! E_e \! -\! E_\nu) \frac{V d\bm p_e}{(2\pi)^3} \frac{V d\bm p_\nu}{(2\pi)^3} \,,
\end{align}
where $E_i$, $E_f$, $E_e$ and $E_\nu$ are respectively the energies of the initial Tritium, of the final Helium, of the electron and of the anti-neutrino. The sum is over all possible final states for the Helium nucleus. The matrix element for a given final state simply reads, 
\begin{align} \label{eq:Mf}
    \mathcal{M}_f = \frac{g}{V} \int d\bm x \, \psi_{i,\rm T}(\bm x) \, \psi_{f,\rm He}^*(\bm x) \, e^{-i (\bm p_e + \bm p_\nu) \cdot \bm x} \,.
\end{align}
It is then clear that to determine the $\beta$-decay rate one must know the wave function of the initial Tritium, as well as the wave functions of the possible final states of Helium. In turns, this requires an understanding of how both of them interact with the underlying graphene, before and after the decay has happened. The work presented in the following sections is devoted precisely to this. In particular, we highlight that in earlier studies~\cite{PTOLEMY:2022ldz} it was essentially assumed that the interactions of Tritium and Helium with graphene were essentially the same. Our work revises this assumption, by studying how the graphene systems responds to the change in the nuclear charge. We will get back to the decay rate, and its evaluation, in Section~\ref{sec:finalrate}.

Before moving on, let us stress right away that the rate presented in Eqs.~\eqref{eq:dgamma} and \eqref{eq:Mf} cannot be considered as the end of the story. In fact, it does not account for the possible initial and final state contributions coming from, for example, the electronic and vibrational degrees of freedom characterizing the graphene~\cite[e.g.,][]{Tan:2022eke}. We will comment more on this in Section~\ref{sec:limitations}. 
\section{Tritium before the decay} \label{sec:initial}
\begin{figure*} 
\includegraphics[width=0.85\linewidth]{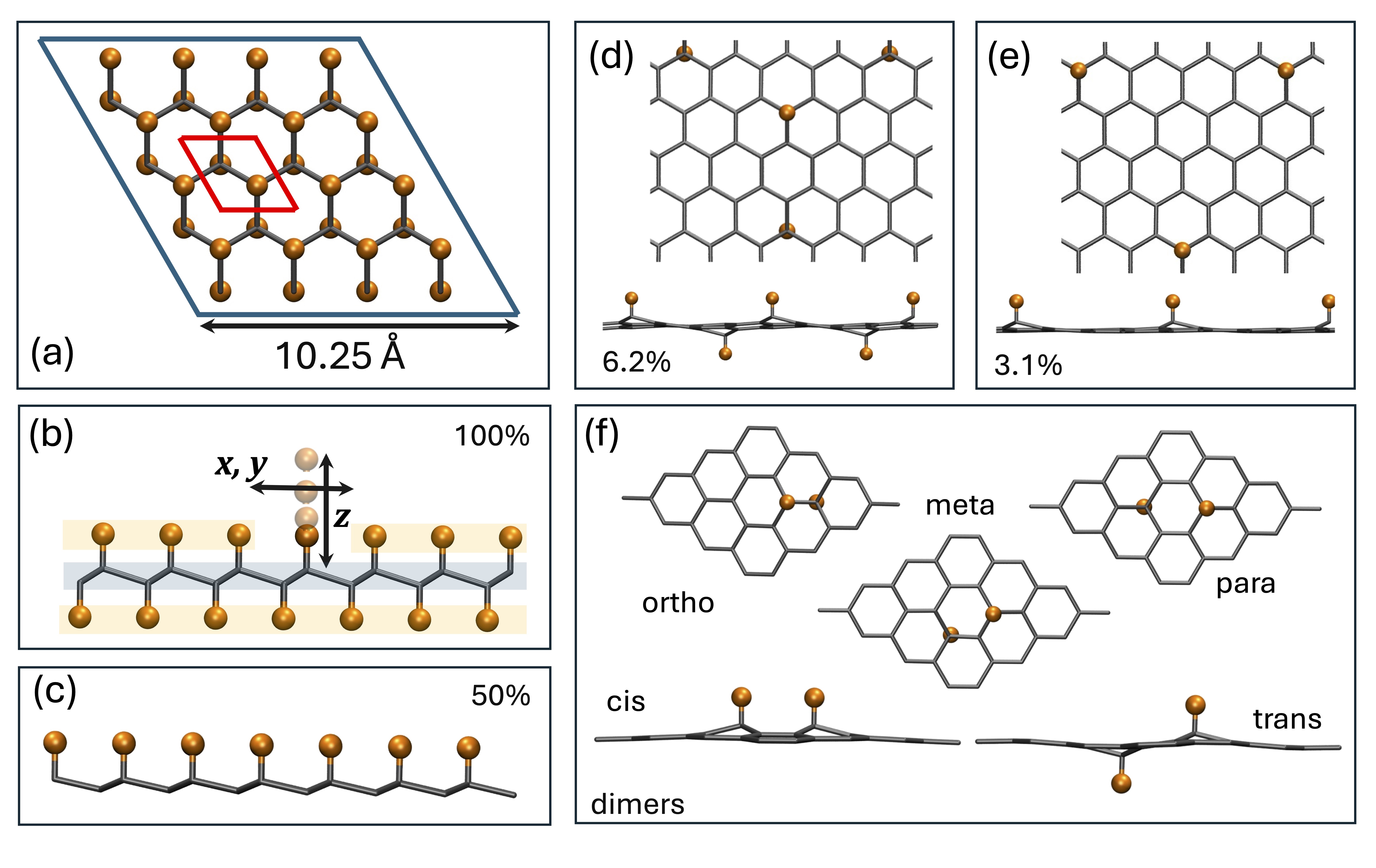}
\caption{Model system and calculation setup. The structures include 32 C atoms and up to 32 T atoms and are extended using periodic boundary conditions. The supercell (whose boundaries are in blue in (a)) is the 4$\times$4 repeat of the unit cell (in red in (a)). The periodicity in the $z$-direction is 22~\AA. The exchange and correlation functional is the Perdew-Burke-Ernzerhof~\cite{PBE} parametrization of the Generalized Gradient Approximation. The core electrons are treated with the Projector Augmented Wave approach and with pseudopotentials~\cite{Lejaeghere_2016}, and the cutoff for the plane wave expansion is $k^2/2=65$~Ry. Grimme Van der Waals empirical corrections \cite{Grimme_2006} are included.
The Brillouin zone is sampled with $18\times18\times 1$ symmetrically distributed $k$-points~\cite{Monkhorst_1976}. When structure relaxation is needed (e.g., panel (a) of Fig.~\ref{fig:scheme}), we reduce the forces down to 10$^{-3}$~Ry/Bohr. 
{\bf Panel (a):} Top view of the fully loaded chair structure. {\bf Panel (b):} Side view of the same. The coordinates along which the Tritium is moved are also reported. In the first prescription for the Tritium potential, all other atoms are allowed to relax, in the second one the carbon ones are kept frozen (gray shaded region), while in the third prescription the other Tritium atoms are also kept frozen (orange shaded region).
{\bf Panel (c):} Side view of the 50\% loaded structure. {\bf Panels (d) and (e):} Models of the 6.3\% and 3.1\% loaded systems. The distance between two Tritium atoms is approximately 9.7~\AA \ and 5.6~\AA, respectively. {\bf Panel (f):} Tritium loaded in ``dimers'', in ortho, meta and para configuration, each considered in cis and trans conformations, as defined in the figure itself.}
 \label{fig:models}
 \end{figure*}
\noindent
We start by studying extensively the initial state of Tritium bound to graphene (Fig.~\ref{fig:scheme} (b)), for different levels of loading, spatial distribution and magnetization state (see also Fig.~\ref{fig:models}). 
Upstream of the DFT scheme employed to determine the electronic structure, there is the adiabatic approximation for the electrons~\cite[e.g.,][]{weinberg2013lectures}. Briefly, the very large ratio between the electron and nuclear masses allows to consider the electronic state as adiabatically following the nuclear motion. In other words, one solves for the electronic structure for any given nuclear configuration, defining electronic eigenstate energies which are parametrically dependent on the nuclear position, $E_n(\{\bm X \})$. 
These eventually provide effective potential energy surfaces (PES) for the nuclear dynamics, one for each given electronic state, $n$. This framework --- the so-called Born-Oppenheimer (BO) approximation~\cite{born1988dynamical} --- is valid under the additional condition that the PES corresponding to different electronic states remain well separated in energy. For Hydrogenated graphene, the electronic gap between the ground and first excited states has been shown to increase with the amount of hydrogenation, up to $5\text{-}6$~eV~\cite{rossi15,betti22,apponi2025highlyhydrogenatedmonolayergraphene}. This allows us to safely use the BO approximation for the ground state. 

In our case, the ground state PES depends on the coordinates of the Tritium and Carbon nuclei present in the system, $E_0(\{\bm X_{\rm T}\}, \{ \bm X_{\rm C} \})$. For our purposes, we are interested in the potential felt by a single Tritium nucleus, as a functions of its coordinate. We define this as,
\begin{align} \label{eq:VT}
    \! U_{\rm T}(\bm x) \equiv E_0\left(\bm x = \bm X_{\rm T,1},\bar{\bm X}_{\rm T,2}, \dots, \bar{\bm X}_{\rm C,1}. \bar{\bm X}_{\rm C,2}, \dots \right) \, .
\end{align}
Given that $E_0$ is a many-body energy function, the definition above must be supplemented with a prescription on how to treat the other nuclear coordinates, $\{ \bar{\bm X} \}$. Specifically, we consider three possible prescriptions:
\begin{itemize}
    \item For each value of $\bm x$, we allow the other coordinates to fully relax to their values corresponding to the lowest energy configuration. This, however, makes the result very dependent on the environmental factors, such as possible additional restrains that limit the extent of relaxation.
    \item Since the typical relaxation times are controlled by the ratio $(M_{\rm T}/M_{\rm C})$, this implies that the Carbon scaffold  (shaded in gray in Fig.~\ref{fig:models} (b)) relaxes more slowly than the Tritium. On short time scales, thus, the Carbon nuclei might be kept fixed allowing only the Tritium ones to relax.
    \item When considering time scales shorter than the vibrational period of the C-T bond (tens of femtoseconds), one is allowed to only vary the coordinate of the Tritium of interest, with all the other nuclei kept frozen. This option has the advantage of being independent on environmental factors.
\end{itemize} 
In the rest of this work we considered the latter option as the preferred one, although, in some specific cases, all three options are evaluated and compared.  Additional details are reported in the Supporting Information (SI), Section~{\color{Maroon} SI.I}\cite{supplementary}.
\subsection{DFT calculations setup and model systems} \label{sec:model}
\noindent Within the DFT scheme, the electronic structure and ground state energy is calculated solving the Schr\"odinger equation for single electrons, within a self-consistent potential depending on the electronic density, the so-called Kohn-Sham scheme~\cite{giustino2014materials}. For our purposes, it is useful to separate the ground state energy in three terms, with different dependencies on the electron density and nuclear positions, namely,
\begin{align} \label{eq:dft}
   \begin{split}
        &E_0[n,\{{\bm X}\}] = 
        E^{\rm int}[n] + E^{e{\rm N}}[n,\{\bm X\}] +E^{\rm NN} (\{ \bm X \}) \,,
    \end{split}
\end{align}
whose explicit expressions are given in the the SI, Section~{\color{Maroon} SI.I}\cite{supplementary}. Here $E^{\rm int}$ includes the electron kinetic energy and electron-electron interactions, and is a functional of the ground state electronic density $n(\bm x)$ only. Moreover, $E^{e{\rm N}}$ is the direct Coulomb integral between the electronic charge and the nuclear charges, explicitly depending on both $n$ and $\{\bm X\}$, while $E^{\rm NN}$ is the interaction among nuclei and depends only on the nuclear coordinates. Within the standard DFT framework, $n(\bm x)$ is determined self-consistently for any given nuclear configuration, and therefore depends implicitly on $\{\bm X\}$. 

The practical implementation of the DFT scheme requires to specify the functional forms of $E_0[n,\{ \bm X\}]$, which are known for $E^{e\rm N}$ and $E^{\rm NN}$, but not entirely for $E^{\rm int}$: The {\it exchange and correlation} functional, i.e., the part of electron-electron interaction exceeding the direct Coulomb (Hartree) term, is not exactly known, although several approximate forms are available. Here we choose the Generalize Gradient Approximation in the parametric form of Perdew-Burke-Ernzerhof~\cite{PBE}, well tested on similar systems~\cite{rossi15,cavallucci19,PTOLEMY:2022ldz} and displaying high accuracy on hydrogen dissociation\cite{Yu16}. 
 
The potential $U_{\rm T}$ defined in Eq.~\eqref{eq:VT} has been computed for the different levels of Tritium loading, obtained by spatially distributing it as illustrated in Fig.~\ref{fig:models}. For the cases of 100\% and 50\% loading (panels (a)-(b) and (c), respectively), we chose symmetric distributions, such that alternate sub-lattices are occupied on opposite sides (for the 100\% case, also called ``chair'' conformation) or on a single side (for the 50\% case). These conformations, although among the most stable ones, 
are considered as representative ideal reference cases. In real systems, more disordered distributions might be entropically favored, especially when ripples or defects are already present~\cite{delfino_24,goler13}, which tend to clusterize hydrogenation. While here we are forced to regular distributions by the periodic boundary conditions, we note that the binding energy depends on the local occupancy around the site under consideration~\cite{rossi15}. Thus, these distributions may be considered representative also of a lower level of loading in the presence of clusters. Much lower levels of loading are here represented arranging fewer Tritium atoms in evenly spaced positions (panel (d) and (e) for 6.2\% and 3.1\%, respectively), or in ``dimers'', i.e. adjacent locations (panel (f)). In most instances, the calculations are performed without restriction on the electronic spin state, in order to detect possible magnetic states, while in some cases the spin has been restricted to a direction orthogonal to the graphene sheet, to emulate the presence of an external magnetic field. 

Calculations were performed with Quantum Espresso v7.2 \cite{QE-2017}. Input preparation and output analysis were performed with Xcrysden~\cite{XCrySDen} and VMD~\cite{vmd}.
\subsection{The Tritium orthogonal potential} \label{sec:VTortho}
\begin{figure*}[ht]
\includegraphics[width=0.98\linewidth]{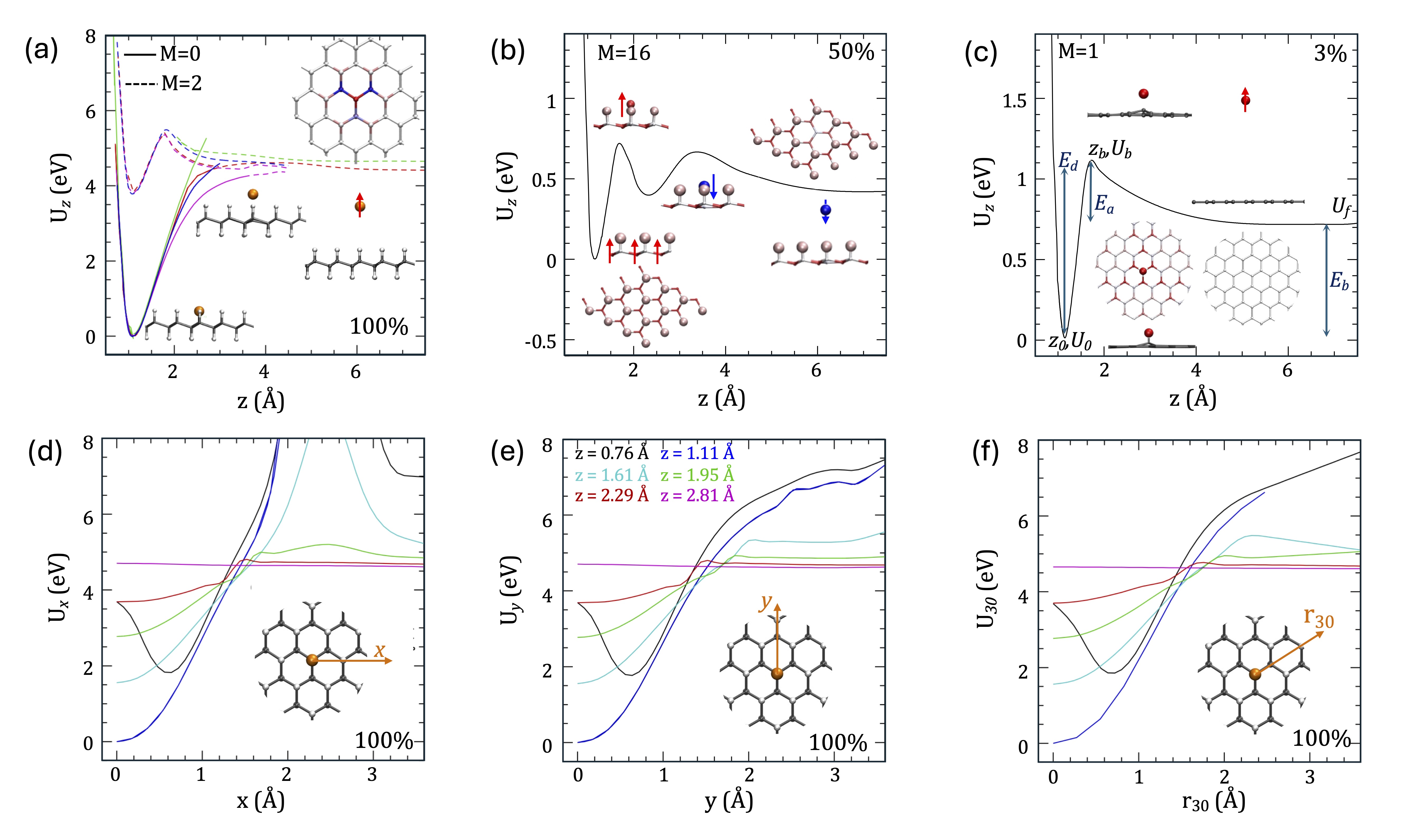}
\caption{Sample orthogonal and parallel potentials.  {\bf Panel (a):} $U_z$ at 100\% loading. Solid and dashed lines correspond to different magnetization of the system, as indicated in the legend, while different colors correspond to different conditions for the relaxation of nuclei. Specifically: red and green are obtained slowly dragging the Tritium under consideration, and relaxing the rest of the system during dragging, either keeping fixed only the Carbon nucleus right below (red), or all Carbon nuclei (green). Magenta lines are obtained fixing the C-T distance at evenly separated values along the path of the Tritium and relaxing all the rest, while for the blue line all Carbons are kept fixed in their starting position (more details on the calculation are reported in~{\color{Maroon}SI.III.A}). Side views of the structures in the minimum, barrier and detached states are reported under the plots. A top view of the system (unbound state) is also reported, colored according to the local magnetization  (red for spin up, blue for spin down).  
{\bf Panel (b):} $U_z$  computed at 50\% loading with symmetrically distributed Tritium, evaluated with free magnetization, which turns out to be $M=16\mu_{\rm B}$ ($1\mu_{\rm B}$ per unit cell), and with the full relaxation prescription. The structures, located near their corresponding place along the curve, illustrate the distribution of magnetization, with coloring as in panel (a), and the arrows indicating the direction of the spin. {\bf Panel (c):} $U_z$ for the case with lowest loading. Side and top views of the structures are also reported, colored according to magnetization as in the previous panels. We also show the definitions of the various energies and special positions discussed in the main text.
{\bf Panels (d)-(f):} Parallel potentials for $100\%$ loading, with as a function $x$, $y$, and a direction at $30^\circ$ from the horizontal one, as indicated in the structures shown in each panel. The different colors correspond to different values of the $z$-coordinate, fixed at different heights, as indicated in the legend of panel (e). The potentials are evaluated keeping fixed all nuclei except the Tritium under consideration.}
 \label{fig:zpotentials}
 \end{figure*}

\noindent For each of the prescriptions described in Section~\ref{sec:initial}, 
we evaluate the Tritium potential as a function the $z$-coordinate, perpendicular to the sheet, with $x$ and $y$, the coordinates parallel to it,  which are kept fixed, 
\begin{align}
    U_{z}(z) \equiv U_{\rm T}(x=0\,, y=0\,, z) \,.
\end{align}
In most cases, the qualitative shape of the potential is similar, and exemplified by the potential of extraction of an isolated Tritium (Fig.~\ref{fig:zpotentials}, panel (c)). It displays a minimum with energy $U_0$, at $z_0\simeq 1.1$~\AA,\footnote{The origin, $z=0$, is taken as the location of the Carbon nucleus to which Tritium is bound, so that $z$ is the length of the T-C bond.} a deep rise for $z < z_0$ (i.e., when the the Tritium gets closer to the Carbon), and a softer rise for $z>z_0$, culminating in a maximum of energy $U_b$, located at $z_b$. This defines a desorption energy barrier, $E_d \equiv U_b-U_0$. After the barrier, at large $z$, the energy may decrease to a final value, $U_f$, defining a barrier for the inverse adsorption process, $E_a \equiv U_b-U_f$. The depth of the binding potential is, instead, defined as the difference $E_b \equiv U_0-U_f = E_a-E_d$, and is negative if bound states are allowed (see again Fig.~\ref{fig:zpotentials}, panel (c)). Given this qualitative behavior, however, the value of the barriers depends on the level of loading and on the spatial distribution of Tritium, as well as on the magnetization state of the system. In some instances, there might even be multiple barriers, or none at all. The energies $E_b$, $E_d$ and $E_a$, as well as other parameters, are reported in the supporting Table~{\color{Maroon} SI.1}\cite{supplementary}, for all cases studied.  

Let us now describe some specific cases more in detail. Panel (a) of Fig.~\ref{fig:zpotentials} reports the orthogonal potential, $U_z$, evaluated at 100\% loading. Looking at the binding minimum, one can see two main groups of curves, having low and high values of $U_0$, separated by $\approx \! 4$~eV. The lower energy curves are obtained leaving the magnetization of the system free to adjust spontaneously, while the higher energy curves are obtained restraining it to $M=2\mu_{\rm B}$, where $\mu_{\rm B}$ is the Bohr magneton. When the magnetization is free, the preferred one for the Tritium bound state is $M=0$ (spin singlet, the solid part of the curves). However, as Tritium is extracted, it drags its single electron away from the substrate, preventing it from coupling with another one to form a singlet state. Therefore, when this happens, the system displays a magnetization $M=2\mu_{\rm B}$ (dashed part of the curves), which is due both to the spin of the electron pertaining to the detached Tritium, and to the spin of the dangling bond. The spin density of the latter distributes around the vacancy on the substrate, as shown by the top view of the structure reported in Fig.~\ref{fig:zpotentials}, panel (a) (red sites = spin up, blue sites = spin down). If, on the other hand, the magnetization is forced to be $M=2\mu_{\rm B}$ along the whole path (all dashed curves), the energy of the binding minimum increases by $\approx \! 4$~eV. This also implies a decrease in the depth of the potential as compared to the case of free magnetization, from $|E_b| \simeq 4.3$~eV to $|E_b| \simeq 0.5$~eV. Similarly, the desorption energy decreases from $E_d\simeq 4.3$~eV to $E_d\simeq 1.5$~eV (see also Table~{\color{Maroon} SI.1}\cite{supplementary}).

Within the two main groups of curves, there are minor differences concerning the location and height of the barriers. Specifically, allowing for an increasing amount of relaxation of the other nuclei, produces a lowering of the energy profiles in the region $2 \text{ \AA} \lesssim z \lesssim 4 \text{ \AA}$, as well as a disappearance of the small adsorption barrier (see the difference between differently colored curves, as detailed in the caption, and Table~{\color{Maroon} SI.1}\cite{supplementary}  for the numerical values). 

Although the ground state magnetization is generally null, there are some  exceptions. These happen when the atoms are spatially distributed in particularly symmetric configurations, as dictated by the symmetries of the graphene lattice. This can be seen as the union of two triangular sub-lattices. In the case of 100\% loading, half of the Tritium atoms are bound on one side of the sheet, on one of the sub-lattices, and the other half on the other side, and on the other sub-lattice (see Fig.~\ref{fig:models}, panel~(a)). When the system is magnetized, spins on adjacent sites tend to have opposite directions and therefore same direction spin tend to occupy alternate sites, i.e. to be on the same sub-lattice. This is shown in the top view of the structure reported in Fig.~\ref{fig:zpotentials}, panel (c). The Tritium generates magnetization on the sheet on alternate sites around its own (sites with spin up density are colored in red), and similar situations occur when the symmetry of the bound Tritium atoms is favorable (see structures reported in Table~{\color{Maroon} SI.1}\cite{supplementary}). 

The case of 50\% loading we consider here, with Tritium loaded on a single side and on alternate sites, is especially favorable to magnetization, since all the sites of a given sub-lattice are occupied and all those of the other are empty. Because of this, the unpaired electrons are all on the same sub-lattice and tend to align their spins, producing a macroscopic magnetization ($M=\pm 1\mu_{\rm B}$ for each unit cell), with a gain in energy with respect to the non-magnetic state of $\simeq 0.23$~eV/unit cell.
On the other hand, the Tritium binding is looser than in the fully loaded case (see Fig.~\ref{fig:zpotentials}, panel (b)). Furthermore, when the Tritium detaches and carries away an unpaired electron with itself, spin rearrangements are needed, which produces multiple barriers. The barrier observed for the detachment of an isolated Tritium at low coverage (Fig.~\ref{fig:zpotentials}, panel (c)) is due to similar spin rearrangements: in the bound state, the additional electron induces a spin density distribution in alternate sites around Tritium, as previously noted. Upon detachment, the graphene recovers its spin neutrality.

In summary, there is a marked dependence of the binding and desorption energies on the Tritium coverage and distribution, which act both directly, modifying the local structure, and indirectly, producing a spontaneous magnetization in given configurations, which in turn alters the potential profiles. Overall, the values of the desorption energy, $E_{d}$, 
range roughly from $0.7$~eV to $5.3$~eV. The binding energy, $E_{b}$, has a similar spread, in agreement with the literature~\cite{delfino_24}. However, we observe that the magnetic configuration may be additionally stabilized/destabilized by the presence of external magnetic fields, as those envisioned in PTOLEMY's setup.
\subsection{The Tritium parallel potential} \label{sec:VTparal}
\noindent To further probe the interaction between Tritium and graphene we also studied how the potential changes when moving parallelly to the graphene plane, exploring it along three symmetry directions, namely,
\begin{subequations} \label{eq:Vpar}
    \begin{align}
        U_x(x\,,z) \equiv{}& U_{\rm T}(x \,, y=0\, ,z) \,, \\
        U_y(y\,,z) \equiv{}& U_{\rm T}(x=0 \,,y\,,z) \,, \\
        U_{30}(r_{30}\,,z) \equiv{}& U_{\rm T}\!\left(x=\tfrac{\sqrt{3}}{2} \, r_{30} \,, y=\tfrac{1}{2} \, r_{30} \,,z\right) \,,
    \end{align}
\end{subequations}
while the $z$-coordinate is kept fixed at different values. The results are reported in Fig.~\ref{fig:zpotentials}, panels (d)-(f), where the symmetry directions are explicitly shown, and the potentials corresponding to different values of the $z$-coordinate are plotted in different colors. 

Importantly, in the region of about {$1.5$~\AA} around the equilibrium point, the potential is very similar in all three directions, due to the high degree of isotropy induced by the hexagonal symmetry of the graphene lattice. In this region, the curves corresponding to $z<z_0$ (black line) display repulsion at $x=y=0$, since the C-T bond is compressed. Conversely, all the other curves display an attractive minimum, which becomes shallower as the distance between Tritium and graphene increases, as expected. 

For displacements larger than {$1.5$~\AA}, instead, large anisotropies appear. In particular, $U_x$ presents a large potential barrier (see, again, Fig.~\ref{fig:zpotentials}, panel (d)), since moving in that direction the Tritium encounters another one bound in meta position at {$x=2.5$~\AA}. In the case of $U_y$ and $U_{30}$, instead, the nearest Tritium is located at {$4.3$~\AA}, and therefore the barrier is considerably farther away, out of the range probed here. These results, together with the potential $U_z$ described in the previous section, give a good description of the complete potential, $U_{\rm T}(\bm x)$, for $100\%$ loading. This will serve as the basis to compute the initial state wave function to be used in the $\beta$-decay rate, as anticipated in Section~\ref{sec:betaintro} and detailed in Section~\ref{sec:finalrate}.
\section{Helium right after the decay}
\label{sec:final}
\noindent
We now move to the analysis of the final state of the reaction in Eq.~\eqref{eq:decay}: Helium  in interaction with the rest of the substrate immediately after the decay.  Near the end-point, the $\beta$-decay happens over a time scale of about $t_\beta \approx 10^{-18}$~s, which is conservatively estimated as the time it takes for the $\beta$-electron, with momentum $p_\beta \approx 100$~keV, to escape the Helium atom, whose radius is $r_{\rm He} \approx 1$~\AA.

Immediately after the decay, the electronic structure will still be the one corresponding to the ground state of the system before the decay. However, for the new system, which now has a Helium nucleus, this will not correspond to the ground state anymore, but it will rather be a combination of excited electronic states (named He$^*$ in panel (c) of Fig.~\ref{fig:scheme}). The time it takes for this state to relax down to the ground state is very variable, depending on the specific mechanisms of non-adiabatic relaxation, but it is generally not smaller than
$t_{\rm r} \approx10^{-16}-10^{-14}$~s~\cite{marciniak19,tichauer22}. We can then safely assume $t_\beta \ll t_{\rm r}$. This circumstance evidently prevents the use of the BO approximation, which requires that the electronic structure corresponds to an eigenstate of the current nuclear configuration. 

In this context, one often uses the so-called sudden approximation~\cite[e.g.,][]{Sakurai:2011zz,Saenz:1997zz}. In our case,  such an approximation is valid because, on the time scale relevant for the decay, the electronic structure of the system does not have enough time to relax, thus remaining the same as it was right before the transition. However, the implementation of this approximation in our context requires to abandon the standard DFT scheme. Specifically, referring to Eq.~\eqref{eq:dft}, the electronic density, $n(\bm x)$, should not be calculated self-consistently given the new nuclear configuration (with a Helium), but must be kept the same as it was before the decay. As a consequence, the term $E^{\rm int}[n]$ in Eq.~\eqref{eq:dft} is the same as the one evaluated before decay. Conversely, the terms $E^{\rm NN}(\{ \bm X \})$ and $E^{e{\rm N}}[n,\{ \bm X \}]$ do vary, since the nuclear charge has now changed. The calculation of these non-self-consistent terms are not implemented in the standard DFT, and therefore require specific post-processing tasks, which we describe in the SI, Section~{\color{Maroon} SI.I.E}\cite{supplementary}.

\begin{figure}
\includegraphics[width=0.85\linewidth]{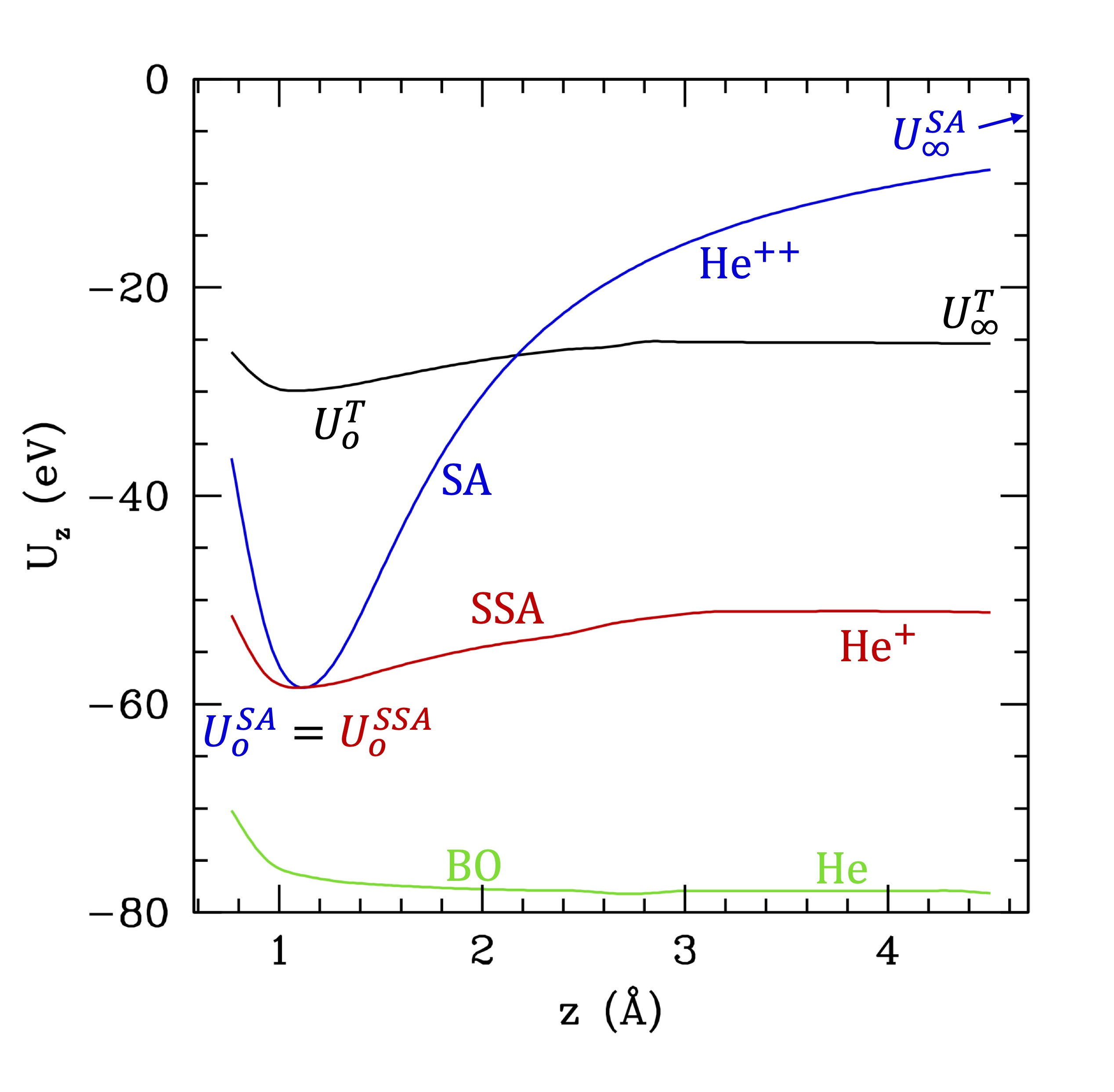}
\caption{Orthogonal Helium potential in sudden and semi-sudden approximation (respectively, SA and SSA, in blue and red), and in the adiabatic (BO) approximation (green). The system is considered electrically isolated during decay, and therefore the total charge is +1, although in the three cases the Helium is extracted in different ionic states, as indicated. The corresponding parallel potentials are reported in Fig.~{\color{Maroon} SI.4}\cite{supplementary}. For comparison, the Tritium potential is also reported, aligned to a common energy frame as explained in the text and in the SI, section {\color{Maroon} SI.I.F}\cite{supplementary}. }\label{fig:potsudden}
\end{figure}

Even within the sudden approximation, the final Helium potential depends on the choice we make for the electronic density, $n$.\footnote{As before, the  potentials also  depend on how other nuclei are treated. For this case, we use the prescription where all the other are kept forzen.} In particular, we implement two possible choices, whose results are reported in Fig.~\ref{fig:potsudden}, panel (a).
\begin{itemize}
    \item The very {\it sudden approximation}: the electronic density employed to compute the Helium potential is 
    the density evaluated before decay, for a value of the Tritium coordinate equal to the equilibrium one, i.e.,
    \begin{align}
        n(\bm x) \equiv n_{\bm X_{\rm T} = (0,0,z_0)}(\bm x) \,.
    \end{align}
    Since in this case the electronic density is independent of the position of the Helium nucleus, and mostly dense near the original equilibrium position, at large distance the Helium is extracted as naked nucleus, He$^{++}$. The corresponding potential is strongly attractive as a consequence of the Coulomb interaction with the negatively charged substrate, as it can be seen in Fig.\ \ref{fig:potsudden}. 
    \item {\it Semi-sudden approximation}: in this case, the electronic density is taken to be the same it was before the decay, but evaluated at a Tritium coordinate equal to the current Helium coordinate,
    \begin{align}
        n(\bm x) \equiv n_{\bm X_{\rm T} = \bm X_{\rm He}}(\bm x) \,.
    \end{align}
    In this case, it turns out that a single electron follows the Helium nucleus. Consequently, for large distances one actually extracts a He$^+$ ion, and its potential is less attractive than in the sudden case.   
\end{itemize}
In Fig.~\ref{fig:potsudden} we also report the Helium potential obtained within the full {\it adiabatic approximation} (i.e., the standard BO framework). In this case, two electrons follow the Helium nucleus, which detaches from the substrate as a neutral atom. As a noble element, an unbound Helium atom is very stable, and therefore its potential is repulsive. The vertical displacement of the adiabatic curve with respect to the sudden/semi-sudden at their minima is due to the electronic structure relaxation, and turns out to be approximately $18$~eV. 
The figure also includes the adiabatic potential for Tritium. Within DFT calculations, the absolute energy reference of systems with different constituents is generally different. Consequently, to correctly align the Tritium potential to the others,  we used as reference the Coulomb tails of potentials for the Helium and Tritium nuclei. These are aligned using the respective sudden potentials (that for Tritium is not shown here, but has been computed for this purpose). More details are given in the SI, section {\color{Maroon} SI.I.F}\cite{supplementary}. This sets the energy difference between the minimum of the Tritium potential, and the minimum of the sudden/semi-sudden potentials, to be $U_0^{\rm T}-U_0^{\rm SA/SSA} \simeq 29$~eV. 

Which of these potentials one should use, is a matter of time scales. As mentioned, the sudden approximation assumes for the electronic density to be the same as it was when the Tritium occupied its equilibrium position, before the decay. However, at least from a semi-classical viewpoint, the Helium nucleus employs some finite time to probe larger distances. If this time exceeds $t_{\rm r}$, the electronic structures may have time to relax, consequently changing the potential felt by the Helium itself. Moreover, our semi-sudden approximation is inspired by a semi-classical picture where the Tritium is slowing moving around, with its motion followed adiabatically by the corresponding electronic density. When it suddenly decays, at a given position $\bm X_{\rm T}$, the electronic density remains frozen to the configuration it had for that position. Therefore, if $t \ll t_{\rm r}$, the potential is expected to be something in between the sudden and semi-sudden case, with the former providing a better description of the regions near the minimum, and the latter of the regions farther away. For much longer times and distances, instead, the electrons will relax, and the repulsive adiabatic potential will gradually provide a better description. 
\section{The $\beta$-decay rate} \label{sec:finalrate}
\noindent We now have all the ingredients to determine the $\beta$-decay rate following from Eqs.~\eqref{eq:dgamma} and \eqref{eq:Mf}. The wave function of the initial Tritium, $\psi_{i,\rm T}$, is taken to be in the ground state of the initial potential, while the wave function for the final Helium nucleus, $\psi_{f,\rm He}$, is evaluated starting from the potentials determined within each of the three schemes described in Section~\ref{sec:final}: sudden, semi-sudden and adiabatic. As we will see, each of these schemes results in a different shape for the $\beta$-electron spectrum. 

Our starting point to determine the wave functions, both initial and final, is the generic form of the Schr\"odinger equation for the nuclei participating in the decay process,
\begin{align}
    - \frac{1}{2 m} \nabla^2 \psi_A(\bm x) + U(\bm x) \psi_A(\bm x) = \varepsilon_A \psi_A(\bm x) \,,
\end{align}
where $m$ is either the Tritium or the Helium nuclear mass, and $A$ is a set of quantum numbers. Depending on whether one is solving for the initial or final wave function, the potential $U$ is either of the potentials described in the previous sections.

\begin{figure*}[th]
    \centering
    \includegraphics[width=\textwidth]{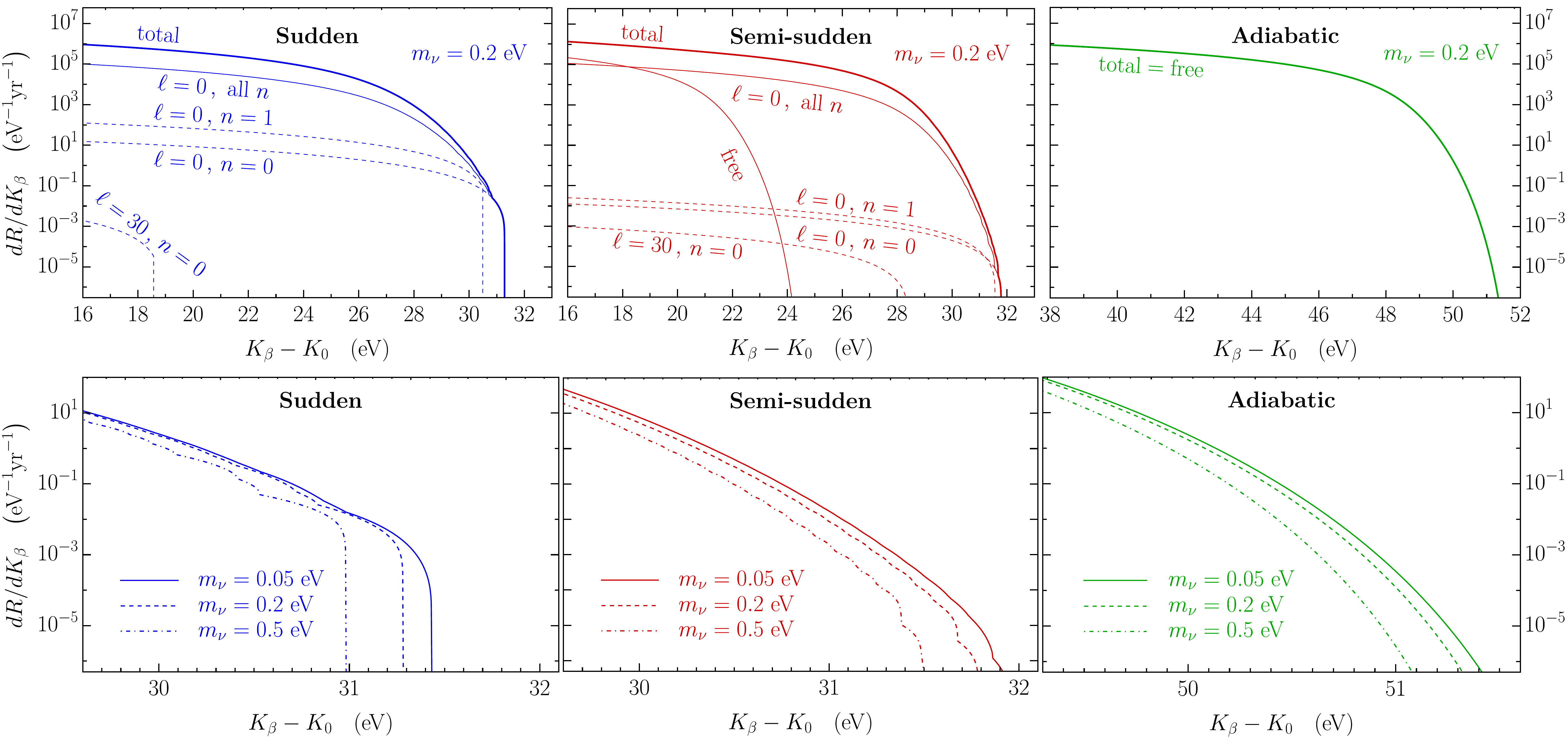}
    \caption{Electron $\beta$-decay spectra obtained for each of the three schemes employed to define the final Helium potential. The initial Tritium potential is taken to be the one corresponding to 100\% loading, computed in the prescription where all other nuclei are kept fixed (see Section~\ref{sec:initial}). The total Tritium mass is taken to be $M_{\rm tot} = 1 \text{ } \upmu\text{g}$. The electron's kinetic energy is measured with respect to the end-point of a Tritium nucleus in vacuum and for a massless neutrino, $K_0$. {\bf Upper panels:} Total rate for $m_\nu = 0.2$~eV. We also show the partial contribution for some selected final states (or groups of final states), as indicated directly in the figures. {\bf Lower panels:} Total rates in a region much closer to the end-point, as computed for different values of the neutrino mass.}
    \label{fig:spectra}
\end{figure*}

We first recall that in the bound region, i.e. for energies $\varepsilon_A < U_f$, the potentials reported in Fig.~\ref{fig:zpotentials} (and Fig.~{\color{Maroon} SI.4}\cite{supplementary})  look approximately the same if computed as a function of the three parallel directions we probed. Large anisotropies only appear for distances sufficiently away from the minimum, as commented in Section \ref{sec:VTparal}. These regions however, give a minor contribution to the matrix element in Eq.~\eqref{eq:Mf}, since the initial Tritium wave function is exponentially localized in the vicinity of the minimum. This allows to approximate the Tritium and Helium potentials as isotropic on the graphene plane, $U(\bm x) \simeq U(r,z)$.
If one is considering a bound state, the dependence of its wave function on the polar angle defined on the graphene plane, $\theta$, is then completely determined to be,
\begin{align}
    \psi_{n \ell}(\bm x) = \chi_{n \ell}(r,z) \, \frac{e^{i \ell \theta}}{\sqrt{2\pi}} \,,
\end{align}
with $\ell \in \mathds{Z}$. The quantum number $n$ labels the bound state solution to the remaining Schr\"odinger equation,
\begin{align} \label{eq:schro_nl}
    \begin{split}
        - \frac{1}{2m} & \left[ \frac{1}{r} \frac{\partial}{\partial r} \left( r \frac{\partial}{\partial r} \right) - \frac{\ell^2}{r^2} + \frac{\partial^2}{\partial z^2} \right] \chi_{n \ell} \\
        & \qquad\qquad \qquad + U(r,z) \, \chi_{n\ell} = \varepsilon_{n \ell} \, \chi_{n \ell} \,.
    \end{split}
\end{align}
The initial state of Tritium is assumed to be the ground state of the potential described in Sections~\ref{sec:VTortho} and \ref{sec:VTparal}, so that the wave functions describing the Tritium and a generic Helium bound state are,
\begin{align}
    \psi_{i,\mathrm{T}}(\bm x) = \frac{\chi_{00}^{\mathrm{T}}(r,z)}{\sqrt{2\pi}} \,, \quad \psi_{f,\mathrm{He}}(\bm x) = \chi_{n \ell}^{\mathrm{He}}(r,z) \, \frac{e^{i \ell \theta}}{\sqrt{2\pi}} \, .
\end{align}
For final states where the Helium is, instead, in the continuous spectrum (i.e., when it is freed from the graphene right after the decay), we simply approximate the wave function by a plane wave. The details on how we solve the Schr\"odinger equation, and how we compute the event rates, are reported extensively in the SI, Section~{\color{Maroon} SI.II}\cite{supplementary}.

In Fig.~\ref{fig:spectra}, we report the resulting electron spectra, as obtained for a total Tritium mass of $M_{\rm tot} = 1 \text{ } \upmu\text{g}$, corresponding to about $2 \times 10^{17}$ Tritium atoms. For the sake of this work, we consider the simpler instance of a single neutrino species, with mass $m_\nu$. In the upper panel we give an extended view of the partial contributions coming from various final states, for $m_\nu = 0.2$~eV. In the lower panel, instead, we zoom in on the end-point region, and show how the expected rate changes with varying neutrino mass.

As one can see, the different schemes for the determination of the final Helium potential produce $\beta$-decay spectra that share some similarities, but also some differences. In particular, both the spectra obtained within the sudden and semi-sudden schemes show several features in the near-end-point region, each associated to the contribution due to a different bound state that the Helium nucleus can end up in. These are very characteristic of the system that hosts the decaying nucleus, thus differentiating the rate expected for a setup like KATRIN~\cite{Aker:2024drp}, PTOLEMY~\cite{PTOLEMY:2019hkd} or Project8~\cite{Project8:2022wqh}.

Nonetheless, the rate obtained with different schemes, differ in a few ways. First of all, when final Helium bound states are allowed, the maximum electron energy is,
\begin{align}
    \begin{split}
        K_{\beta}^{\mathrm{end}} \equiv{}& Q - m _{\nu} + \varepsilon_{00}^{\rm T} - \varepsilon_{00}^{\rm He} \\
        ={}& Q - m_\nu + U_0^{\rm T} - U_0^{\rm He} \\
        & + \frac{1}{2}\big( \omega_{\rm t}^{\rm T} - \omega_{\rm t}^{\rm He} \big) + \big( \omega_{\rm p}^{\rm T} - \omega_{\rm p}^{\rm He} \big) \,,
        \end{split}
\end{align}
where $Q \equiv m_{\rm T} - m_{\rm He} - m_e$ is the $Q$-value determined by the Tritium and Helium {\it nuclear} masses, and $\varepsilon_{00}^{\rm T}$ and $\varepsilon_{00}^{\rm He}$ are the energies of the ground state of Tritium and Helium, respectively. In the last line we have expressed these energies in terms of the parameters of the potentials, in particular, the value of their minima, $U_0$, the fundamental frequency in the direction transverse to the graphene, $\omega_{\rm t}$, and the fundamental frequency in the direction parallel to the graphene, $\omega_{\rm p}$. Now, while the minimum of the potential is the same for both the sudden and semi-sudden schemes, the fundamental frequencies are not, as they depend on the actual shape of the potential close to its minimum. 

The two schemes also differ in the separation between the end-point and the onset of the contribution coming from the Helium continuous final states. In fact, the energy at which the Helium can be freed from the graphene layer is determined by the difference between the energy of the initial bound Tritium, and the value taken by the Helium potential at infinity, $U_\infty^{\rm He}$. Specifically, it is given by,
\begin{align}
    \begin{split}
        K_\beta^{\rm free} \equiv{}& Q - m_\nu + \varepsilon_{00}^{\rm T} - U_\infty^{\rm He} \\
        ={}& Q - m_\nu + U_0^{\rm T} + \frac{1}{2} \omega_{\rm t}^{\rm T} + \omega_{\rm p}^{\rm T} - U_\infty^{\rm He} \,.
    \end{split}
\end{align}
Since the value of the potential at infinity is much lower for the semi-sudden approximation than it is for sudden one, the onset of the continuum in the former is substantially closer to the end-point than it is for the latter.\footnote{Indeed, for the sudden case, the onset of the continuous states is so far from the end-point that it does not show in our plot.} For energies sufficiently far away from the end-point (but not so far away that the neutrino momentum can no longer be neglected), the contribution coming from continuous final states might dominate the total decay rate, and consequently the experimental statistics.

Finally, the spectra associated to each final bound state (some of them shown as dashed lines) are more widely separated in the sudden case. This is because the potential well is narrower, thus causing a wider separation between energy levels. The latter quantity is the one setting the distance between the various end-points of the partial rates associated to different Helium bound states.

For values of the electron energy sufficiently below the end-point, the features mentioned above are not visible anymore. This is because the various discrete final states get denser and denser, effectively making the difference between each other vanishingly small.

Lastly, in the adiabatic scheme the final Helium potential is purely repulsive and, therefore, does not allow for bound states. Consequently, the electron spectrum gets contributions only from the continuous final states.

Before moving on, we highlight a crucial fact. The PTOLEMY project aims at an energy resolution of about 100~meV. As compared to that, the difference between the end-point energies we are predicting here and the one expected in vacuum is extremely large. This constitutes a clear, experimentally measurable signature of the fact that the presence of a solid state substrate has consequences on the $\beta$-decay spectrum.
\section{The fate of Helium}\label{sec:fate}
\noindent
After the nuclear reaction, when both the $\beta$-electron and the anti-neutrino are far away, the system is left with Helium in place of Tritium, in an excited electronic state, and will tend to relax to the new electronic ground state.  Panel~(d) of Fig.~\ref{fig:scheme} reports a view of the electronic density rearrangement upon this relaxation, determined by comparing the electronic density, $n(\bm x)$, evaluated right before the decay with the Tritium in its equilibrium position, $\bm z_0$, with that evaluated after full electronic relaxation with the Helium also in $\bm z_0$. We find that the electrons deplete the light blue areas and populate the pink ones (a top view of the same figure is reported in the SI Fig.~{\color{Maroon} SI.1})\cite{supplementary}. This is simply understood, as they tend to move towards the Helium nucleus, to partially compensate for its additional positive charge. 

\begin{figure}[th!]
\includegraphics[width=0.85\linewidth]{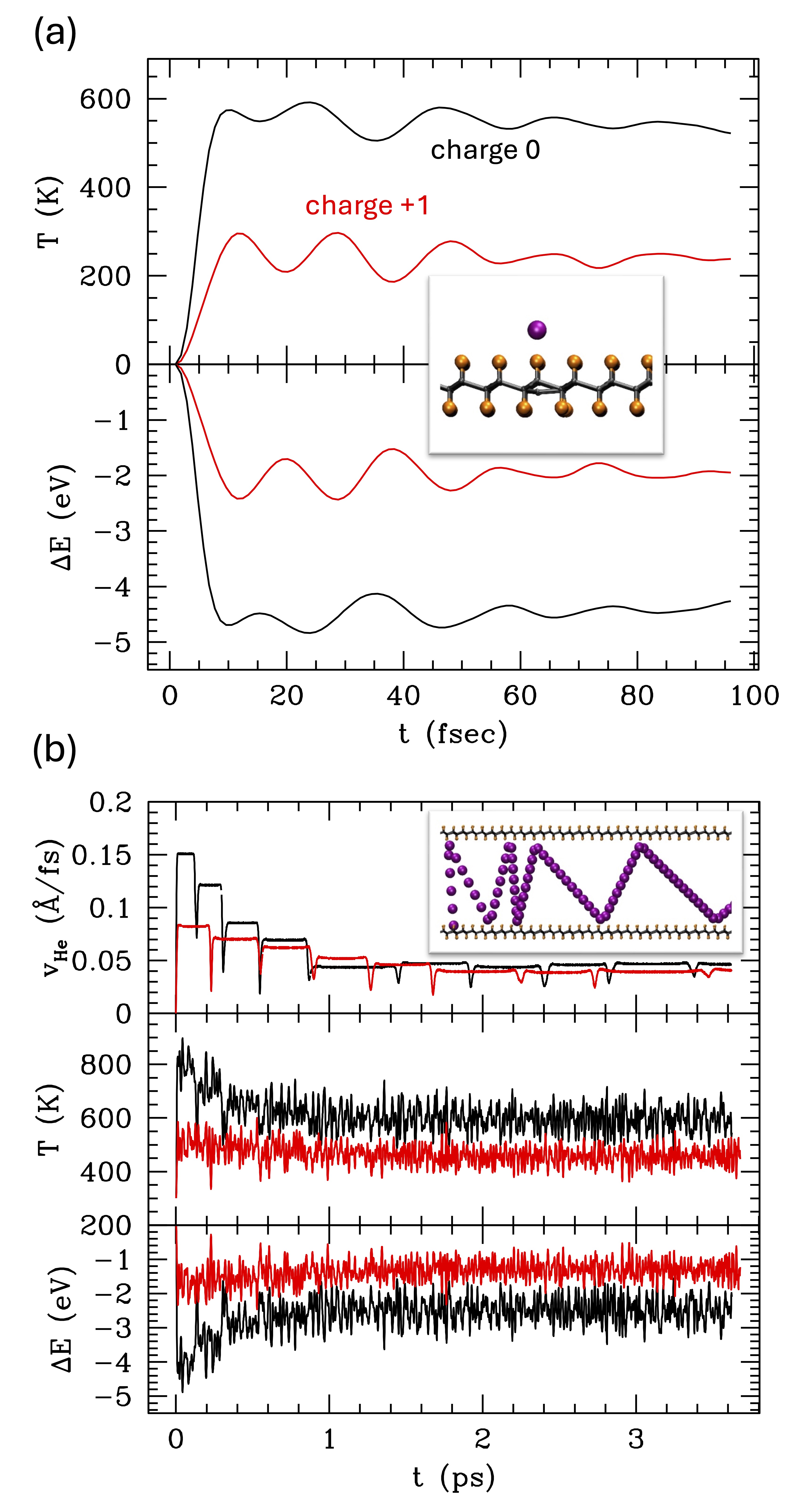}
\caption{Evolution of $\Delta E \equiv E_0(t)-E_0(t=0)$, temperatures and Helium velocities with BO molecular dynamics after the nuclear decay. In both cases the black curves are obtained neutralizing the system with an additional electron to compensate the additional positive charge of Helium, the red ones are instead for an electrically isolated system, which then carries a global charge of +1. Given a certain kinetic energy of the nuclei, $K$, the temperature is defined by $K \equiv \frac{3}{2} N k_{\rm B} T$, with $N=32$ the number of Carbon atoms in the supercell. {\bf Panel~(a):} the starting configuration is the one in equilibrium before decay, and the starting velocity are null. The structure reported as inset shows that the C site where He is detached is deformed by recoil. {\bf Panel~(b):} the starting configuration is the same and the velocities are randomly distributed to set the starting temperature at 300~K. In this case, we also report the Helium velocity, in the top plot. A representation of the Helium trajectory bumping between layers is also reported (Helium in purple). The system is evolved with Newtonian dynamics for the nuclei, integrating with Verlet algorithm and with a time step of $0.97$~fs.}
\label{fig:BO}
\end{figure}
After electronic relaxation, the BO approximation can be safely used again. In the simulations reported here, however, we further assume that the energy that the Helium nucleus has possibly acquired from the nuclear reaction, is small or dissipated during the time $t_{\rm r}$, during which the attractive sudden potential lives (the opposite situation, in which the Helium nucleus is freed right away from the sheet, is considered in \cite{apponi2025stabilityhighlyhydrogenatedmonolayer}.) The potential calculated in the BO approximation  (Fig.~\ref{fig:potsudden}, blue line) pertains to the neutral atom and is repulsive. Consequently, Helium is eventually released from the structure. We explore the release and the following motion of the substrate performing BO molecular dynamics, i.e. solving the classical motion on the ground state PES. 

Fig.~\ref{fig:BO}, panel (a) reports the dynamics of the system for the first picosecond, starting from the equilibrium configuration before the decay with null velocities (i.e., roughly zero temperature). We have studied the dynamics of the system for two different configurations: electronically isolated from the outside, and grounded. The red curves in Fig.~\ref{fig:BO} correspond to the first instance, in which the system has a total charge of +1, while the black curves correspond to the second instance, where the system is neutralized by gaining an electron from the outside. We find that the Helium detaches in few fs with an energy gain for the system $>2$~eV, as expected (numerical values of these energies for different cases, are reported in the SI, Fig.~{\color{Maroon} SI.1}\cite{supplementary}). The oscillations which are visible in Fig.~\ref{fig:BO} are due to the vibrational excitations induced by the recoil of the Carbon site on which the decay has happened, shown in the structure reported as inset. When the system is neutral (black lines), the behavior is similar, but the energy gain is nearly double, as a consequence of the larger stability of neutral system. 

Just after release, the total kinetic energy gets a contribution due to the velocity acquired by the Helium, which is $v_{\rm He} \approx 2.9 \times 10^{-5} \simeq 0.088$~\AA/fs, for the overall charged case, and  $v_{\rm He} \approx 5 \times 10^{-5} \simeq 0.15$~\AA/fs, when the system is electrically neutral. These values correspond to a Helium kinetic energy of about $1.2$~eV and $3.5$~eV, respectively. 
As the system evolves, Helium bumps on the graphene layers, and gradually distributes its energy to the substrate over a time scale of ps. This is shown in the upper part of Fig.~\ref{fig:BO}, panel (b), where we also show a prototypical trajectory in the inset. As a consequence, the average energy of the system increases of a few eV.

While these energies and temperatures might seem very high, one must consider that the simulations are performed with a small supercell of area $\lesssim 1$~nm$^2$, periodically repeated. This means that we are not actually observing a single Helium release, but rather the simultaneous release of periodically spaced Helium atoms, about $1$~nm apart from each other. Each of them is increasing the system energy by a few eVs. As one can see from the rough relaxation time of the curves in Fig.~\ref{fig:BO}, it takes about 1~ps to distribute this energy away to the rest of the system. The total radioactive activity of Tritium is $d\Gamma /dM \simeq 3.6 \times 10^{14} \text{ g}^{-1}\text{s}^{-1}$. For a total mass $M_{\rm tot} = 1~\upmu{\rm g}$, which is the benchmark considered so far, this implies around one decay every 3~ns, leaving time for the system to cool off. 

Nonetheless, we also ran a second set of simulations, were we intentionally put the system in more challenging conditions, to test the possible damaging effects of the released Helium --- Fig.~\ref{fig:BO}, panel (b). In particular, we set the starting velocities to those corresponding to room temperature, instead of starting with $T=0$. Even in this context, we do not observe any damage produced by the Helium to the structure. Actually, the creation of a vacancy in the Carbon substrate, may only happen in very specific cases, when Helium hits some weaker bonds of the graphene scaffold frontally and with kinetic energy exceeding $\approx \! 3$~eV. There weakers bonds are typically located at the boundaries between T-loaded and T-free areas. 
One such case, created on purpose, is illustrated in the SI, Fig.~{\color{Maroon} SI.2}\cite{supplementary}. Overall, our simulations suggest that these events are extremely unlikely. 
\section{Limitations of our approach and theory systematics} \label{sec:limitations}
\noindent As already explained, to successfully constrain the possible values of the neutrino mass it is of utmost importance to provide a reliable theoretical prediction for the electron's $\beta$-decay spectrum. The present work is a first step in this direction, as we connect the 
properties of Tritiated graphene to the shape exhibited by the spectrum. The framework we are currently working in, however, has some limitations, which we now discuss.

First of all, we are not accounting for the other possible degrees of freedom which can be excited in both the initial and final states of the reaction. These can be vibrational modes of the substrate or electronic excitations. 
If present in the initial state, they might contribute to changing the quantum and statistical properties of the decaying Tritium nucleus, while, if present in the final state, they will subtract energy and momentum from the outgoing electron, inducing further distortions in its spectrum.

Secondly, we frame everything in terms of ``potentials'' for the initial Tritium and final Helium nuclei. At this stage, this is done to simplify the treatment, by reducing the problem to the quantum mechanics of a single particle. In order to define a potential like this one, one must however reduce the intrinsically many-body nature of the problem to that of a single body. Nonetheless, the nuclear system we are dealing with does not feature any truly large hierarchy between masses and/or time scales to rigorously justify this separation, as of now (see the following section for possible ways out).
In light of this, we work under different prescriptions for how to deal with the other nuclear coordinates, as described in Section~\ref{sec:initial}. Each of these prescriptions comes with advantages and disadvantages, and lead to final results that can be different from each other. 

However, upstream of the nuclear dynamics there is a strong assumption we made on the separation between electronic and nuclear dynamics, and further treatment of the electronic problem either in the fully adiabatic (BO) approximation or in its opposite, the (semi-)sudden approximation. While the BO approach is reasonably acceptable before decay and after electronic relaxation, 
the critical temporal region just after the decay is characterized by a strongly non-adiabatic electron dynamics. In this work, we adopted the ``sudden" scheme, which is an extreme simplification valid for the very early instants after the $\beta$-decay, leading to the definition of a possible potential for Helium, accounting for the almost instantaneous change of nuclear charge.  Consequently, on top of the standard BO scheme for the calculation of the Helium potential, we implemented two additional schemes, corresponding to different declination of the sudden approximation. While combining these schemes with the quantum treatment of the nuclear state for the calculation of the decay rates is one of the most innovative aspects of our work, the systematic applicability of the sudden approximation, as combined with the concept of potential, should be investigated deeply, possibly within a non-adiabatic coupled electron nuclear approach (see next section). 

As we reported in the previous sections, the implementation of the different prescriptions and schemes, eventually leads to qualitative and quantitative differences in the final electron spectrum. At this stage, these differences should be taken as a first (very conservative) measure of the theory systematics. This can be practically done, for example, by estimating the sensitivity to the neutrino mass starting from each of the three spectra reported in Fig.~\ref{fig:spectra}, and taking their largest different as theory uncertainty.
\section{Conclusions and Perspectives} \label{sec:conclusions}
\noindent
The issues raised in the previous section could all be summarized in the consideration that the established and more commonly used theoretical scheme generally rely on some sort of separation of energy or time scales, which in turn allows to decouple a subset of degrees of freedom from others, facilitating the approach to the problem. In this work, conversely, we aimed at addressing a question which is at the connection point of high and low energy scales, overlapping on different time scales, bridging solid state with nuclear physics. Therefore approaches relying on the separation of degrees of freedom, although they can be used as the starting point, needs to be reconsidered and possibly revised. 
 
Specifically, we assumed that the degrees of freedom of the Tritium and Helium nuclei involved in the decay can be separated from the other degrees of freedom, in order to evaluate the single Tritium/Helium nuclear wave function. However, the quantum treatment of the full set of nuclear coordinates could be possible within the quasi-harmonic approximation~\cite{cupo19}, which would allow to decouple the degrees of freedom mostly involving the Tritium/Helium, while at the same time evaluating the influence of the substrate vibrational dynamics, on them and on the decay spectrum. The anharmonicity analysis can provide the couplings between the degrees of freedom of interest and the others, providing estimates of the correctness of the decoupling assumption and at the same time of the vibrational relaxation times.

However, a stronger assumption we made is about the decoupling of the electronic coordinates from the nuclear ones. We used it in two rather opposite declinations: the adiabatic approximation, used before and long after the decay, assuming total adiabaticity of the electrons with respect to the nuclear motion, and the sudden approximation, assuming on the contrary frozen electronic structure. Between these two extremes, a number of theoretical schemes are currently considered to address the relaxation of electronic dynamics during nuclear motion, especially when the latter involves light elements, where quantum effects are more evident. In addition to the evolution of electronic excited states (e.g.,\ within the Time Dependent DFT schemes~\cite{marques06}) one must also consider the coupled nuclear motion, either within a semi-classical scheme~\cite{li05}, or within the path-integral scheme~\cite{kim97}, if a full quantum accuracy is required. While the latter is still too computationally heavy to address the system of interest, an interesting approach has recently emerged for a quantum perturbative anharmonic treatment with a limited computational cost~\cite{monacelli21}.

We also point out that, in order to further reduce the theoretical uncertainty on the correct form of the interaction between Helium and graphene, it would be of great relevance to know, experimentally, with which probability the final state corresponds to He, ${\rm He}^+$ or ${\rm He}^{++}$. Such a study for the decay of Tritium on graphene is currently unavailable, although a similar measurement has been performed for the case of molecular Tritium~\cite{TRIMS:2020nsv}.

The analysis of the model system also requires some consideration. Here we considered only a few possible loading levels, and with symmetric distributions. In a real system, the situation is rather different. While the loading can be somehow controlled and measured, the local Tritium distribution at given loading cannot. On the other hand, our results (and others' too~\cite{delfino_24,rossi15,cami15}) indicate that the parameters of the potentials are rather dependent on the local level of loading. Therefore, if one wants to achieve a realistic determination of the parameters of the potential (and in turn of the $\beta$-decay spectrum) one should  consider the statistical distribution of possible environments of Tritium/Helium on graphene. 

Such a statistically extensive analysis cannot be performed with DFT approaches, and one must rely on a much less computationally expensive, fully classical treatment, using empirical inter-atomic potentials. While we have previously performed similar preliminary analyses on small graphene flakes~\cite{delfino_24}, or highly disordered nanoporous forms~\cite{bellucci21}, using reactive potentials~\cite{ftenakis22}, we will eventually need a study focusing on the actual form of graphene substrate proposed by PTOLEMY. This could be, for example, a specific nanoporous form at very low density and low level of defects~\cite{betti22}. To this aim, the recently proposed Neural Network Potentials~\cite{picci25} reaching the same accuracy as DFT, with limited computational cost, may offer an alternative interesting solution.

More broadly speaking, a satisfactory formulation of the Tritium $\beta$-decay rate in presence of graphene will have to address a number of points that are still left open by the current analysis. Among these, for example, is the inclusion of long range electrostatic corrections, due to the interaction between the outgoing electron and the rest of the system. These are known to be rather relevant for the decay of Tritium in vacuum, but their relevance in the present context is for now hard to quantify. It will then be crucial to understand how they distort the spectrum, as they could represent important systematic effects. Moreover, any further refinement of the theoretical predictions should be accompanied by a study of the associated theoretical uncertainties.

All the aspects above will eventually make up the ingredients leading to the determination of the $\beta$-electron spectrum, which should be compared to data to constrain the value of the neutrino mass. To this end, it will be crucial to develop the above proposals, and possibly any other that might be relevant, in the most systematic way possible. 

In light of all this, it is clear that this work is just an initial milestone we have set to frame the problem, but many others must follow.
\subsection*{Acknowledgments}
We are grateful to the whole PTOLEMY collaboration for precious discussions and comments on the manuscript. AE is grateful to Jos\'e~Lorenzana for enlightening discussions on the role of the sudden approximation, and for useful suggestions. GM and VT also thank Paolo~Giannozzi for very useful discussions and support in DFT calculations.  The work of AC and AE has been supported by Sapienza University through the SEED PNR 2022 funds. GM and VT acknowledge the allocation of HPC resources on LEONARDO-CINECA trough ISCRA B SETE (n. HP10BSTFDD) and through the ICSC (Centro Nazionale di Ricerca HPC, Big Data and Quantum Computing, grant n. 1491920). This research was also supported by EU under FETPROACT LESGO (Agreement No. 952068). 
\bibliography{biblio.bib}
\end{document}


\title{The $\beta$-decay spectrum of Tritiated graphene: combining nuclear quantum mechanics with Density Functional Theory} 
\author{Andrea~Casale}
\email{andrea.casale@columbia.edu}
\affiliation{Department of Physics, Columbia University, New York, 538
West 120th Street, NY 10027, USA}
\author{Angelo~Esposito}
\email{angelo.esposito@uniroma1.it}
\affiliation{Dipartimento di Fisica, Sapienza Universit\`a di Roma, Piazzale Aldo Moro 2, I-00185 Rome, Italy}
\affiliation{INFN Sezione di Roma, Piazzale Aldo Moro 2, I-00185 Rome, Italy}
\author{Guido~Menichetti}
\email{guido.menichetti@df.unipi.it}
\affiliation{Dipartimento di Fisica dell’Universit\`a di Pisa, Largo Bruno Pontecorvo 3, I-56127 Pisa, Italy}
\affiliation{Fondazione Istituto Italiano di Tecnologia, Center for Nanotechnology Innovation@NEST, Piazza San Silvestro 12, I-56127 Pisa, Italy}
\author{Valentina~Tozzini}
\email{valentina.tozzini@nano.cnr.it}
\affiliation{Istituto Nanoscienze - CNR, Lab NEST-SNS, Piazza San Silvestro 12, I-56127 Pisa, Italy}
\affiliation{INFN Sezione di Pisa, Largo Bruno Pontecorvo 3, I-56127 Pisa, Italy}
\date{\today}
\begin{abstract}
\vspace{1cm}
\centerline{\LARGE SUPPLEMENTARY MATERIAL }
\vspace{1cm}
\end{abstract}
\maketitle
\section{Details of calculation of Tritium/Helium adsorption/desorption potentials} \label{sec:rest}
\noindent The potential energy surfaces, $E_0(\{\bm X\})$, which we evaluate within the BO and DFT framework (see main text, Section~{\color{Maroon} III}), are many-body potentials of the nuclear coordinates. Here we describe some further detail on how we define a single-body potential for Tritium and Helium. 
\subsection{Treatment of the other nuclear coordinates}
\noindent To calculate the potentials for the single Tritium/Helium moving along the given coordinate, we fix the value of its coordinate, $\bm X$, (e.g., before decay, that of the the Tritium of interest), and provide a prescription for how to treat the coordinates of the remaining nuclei, $\{\bar{\bm X}\}$:
\begin{align} \label{eq:VT}
    \! U_{\rm T}(\bm x) \equiv E_0\left(\bm x = \bm X_{\rm T,1},\bar{\bm X}_{\rm T,2}, \dots, \bar{\bm X}_{\rm C,1}, \bar{\bm X}_{\rm C,2}, \dots \right) \,.
\end{align}

There are different possible prescriptions to specify the rest of the nuclear coordinates. Specifically, we will use the following three options:
\begin{itemize}
    \item {\it One C fixed:}
    \begin{figure}[h!]
        \hspace{2.5em}
        \begin{tikzpicture}			
            \draw[thin,gray] (-3.1,0.1) -- (-3.1,0.5);
            \draw[thin,gray] (-2.4,-0.1) -- (-2.4,-0.5);
            \draw[thin,Maroon] (-1.7,0.1) -- (-1.7,0.5);
            \draw[thin,gray] (-1,-0.1) -- (-1,-0.5);

            \draw[thin] (-3.1,0.1) -- (-2.4,-0.1);
            \draw[thin] (-2.4,-0.1) -- (-1.7,0.1);
            \draw[thin] (-1.7,0.1) -- (-1,-0.1);
        
		\node at (-3.1,0.1) [circle,fill,inner sep=2pt]{};
            \node at (-2.4,-0.1) [circle,fill,inner sep=2pt]{};
            \node at (-1.7,0.1) [circle,fill,inner sep=2pt]{};
            \node at (-1,-0.1) [circle,fill,inner sep=2pt]{};

            \node at (-3.1,0.5) [circle,fill,gray,inner sep=1.5pt]{};
            \node at (-2.4,-0.5) [circle,fill,gray,inner sep=1.5pt]{};
            \node at (-1.7,0.5) [circle,fill,Maroon,inner sep=1.5pt]{};
            \node at (-1,-0.5) [circle,fill,gray,inner sep=1.5pt]{};

            \draw[->,gray,thick] (-0.25,0) -- (0.25,0);

            \draw[thin,gray] (1,0) -- (1,0.25);
            \draw[thin,gray] (1.7,0.) -- (1.7,-0.35);
            \draw[thin,gray,Maroon] (2.4,0) -- (2.4,0.85);
            \draw[thin,gray] (3.1,0.) -- (3.1,-0.35);

            \draw[thin] (1,0.0) -- (1.7,-0.0);
            \draw[thin] (1.7,-0.0) -- (2.4,0.1);
            \draw[thin] (2.4,0.1) -- (3.1,-0.0);

            \node at (1,0.0) [circle,fill,inner sep=2pt]{};
            \node at (1.7,-0.0) [circle,fill,inner sep=2pt]{};
            \node at (2.4,0.1) [circle,fill,inner sep=2pt]{};
            \node at (3.1,-0.0) [circle,fill,inner sep=2pt]{};

            \node at (1,0.25) [circle,fill,gray,inner sep=1.5pt]{};
            \node at (1.7,-0.35) [circle,fill,gray,inner sep=1.5pt]{};
            \node at (2.4,0.85) [circle,fill,Maroon,inner sep=1.5pt]{};
            \node at (3.1,-0.35) [circle,fill,gray,inner sep=1.5pt]{};        

            \node at (2.8,-0.7) {fixed};  
            \draw[->,semithick] (2.75,-0.55) -- (2.47,-0.15);
            \draw[->,semithick,Maroon] (2.4,1.) -- (2.4,1.3);
	\end{tikzpicture}
    \end{figure}
    
    The Carbon nucleus right below the Tritium under consideration is kept completely fixed when varying the coordinate of the latter. For each position of the moving Tritium, all other nuclear positions are allowed to relax. This is equivalent to fixing the C-T distance and relax all other coordinates.
    \item {\it All C fixed:}
    \begin{figure}[h!]
        \hspace{2.5em}
        \begin{tikzpicture}			
            \draw[thin,gray] (-3.1,0.1) -- (-3.1,0.5);
            \draw[thin,gray] (-2.4,-0.1) -- (-2.4,-0.5);
            \draw[thin,Maroon] (-1.7,0.1) -- (-1.7,0.5);
            \draw[thin,gray] (-1,-0.1) -- (-1,-0.5);

            \draw[thin] (-3.1,0.1) -- (-2.4,-0.1);
            \draw[thin] (-2.4,-0.1) -- (-1.7,0.1);
            \draw[thin] (-1.7,0.1) -- (-1,-0.1);
        
		\node at (-3.1,0.1) [circle,fill,inner sep=2pt]{};
            \node at (-2.4,-0.1) [circle,fill,inner sep=2pt]{};
            \node at (-1.7,0.1) [circle,fill,inner sep=2pt]{};
            \node at (-1,-0.1) [circle,fill,inner sep=2pt]{};

            \node at (-3.1,0.5) [circle,fill,gray,inner sep=1.5pt]{};
            \node at (-2.4,-0.5) [circle,fill,gray,inner sep=1.5pt]{};
            \node at (-1.7,0.5) [circle,fill,Maroon,inner sep=1.5pt]{};
            \node at (-1,-0.5) [circle,fill,gray,inner sep=1.5pt]{};

            \draw[->,gray,thick] (-0.25,0) -- (0.25,0);

            \draw[thin,gray] (1,0) -- (1,0.35);
            \draw[thin,gray] (1.7,0) -- (1.7,-0.35);
            \draw[thin,gray,Maroon] (2.4,0) -- (2.4,0.85);
            \draw[thin,gray] (3.1,0) -- (3.1,-0.35);

            \draw[thin] (1,0.1) -- (1.7,-0.1);
            \draw[thin] (1.7,-0.1) -- (2.4,0.1);
            \draw[thin] (2.4,0.1) -- (3.1,-0.1);

            \node at (1,0.1) [circle,fill,inner sep=2pt]{};
            \node at (1.7,-0.1) [circle,fill,inner sep=2pt]{};
            \node at (2.4,0.1) [circle,fill,inner sep=2pt]{};
            \node at (3.1,-0.1) [circle,fill,inner sep=2pt]{};

            \node at (1,0.35) [circle,fill,gray,inner sep=1.5pt]{};
            \node at (1.7,-0.35) [circle,fill,gray,inner sep=1.5pt]{};
            \node at (2.4,0.85) [circle,fill,Maroon,inner sep=1.5pt]{};
            \node at (3.1,-0.35) [circle,fill,gray,inner sep=1.5pt]{}; 

            \node at (2.05,-0.85) {all C fixed};
            \draw[->,semithick,Maroon] (2.4,1.) -- (2.4,1.3);
	\end{tikzpicture}
    \end{figure}
    
    All Carbon nuclei are kept fixed at their initial positions. For each position of the Tritium of interested, the other Tritium nuclei are allowed to relax to their new equilibrium positions. This is equivalent to considering the motion of those Tritium nuclei as adiabatic with respect to that of Carbons.
    \item {\it All fixed:}
    \begin{figure}[h!]
        \hspace{2.5em}
        \begin{tikzpicture}			
            \draw[thin,gray] (-3.1,0.1) -- (-3.1,0.5);
            \draw[thin,gray] (-2.4,-0.1) -- (-2.4,-0.5);
            \draw[thin,Maroon] (-1.7,0.1) -- (-1.7,0.5);
            \draw[thin,gray] (-1,-0.1) -- (-1,-0.5);

            \draw[thin] (-3.1,0.1) -- (-2.4,-0.1);
            \draw[thin] (-2.4,-0.1) -- (-1.7,0.1);
            \draw[thin] (-1.7,0.1) -- (-1,-0.1);
        
		\node at (-3.1,0.1) [circle,fill,inner sep=2pt]{};
            \node at (-2.4,-0.1) [circle,fill,inner sep=2pt]{};
            \node at (-1.7,0.1) [circle,fill,inner sep=2pt]{};
            \node at (-1,-0.1) [circle,fill,inner sep=2pt]{};

            \node at (-3.1,0.5) [circle,fill,gray,inner sep=1.5pt]{};
            \node at (-2.4,-0.5) [circle,fill,gray,inner sep=1.5pt]{};
            \node at (-1.7,0.5) [circle,fill,Maroon,inner sep=1.5pt]{};
            \node at (-1,-0.5) [circle,fill,gray,inner sep=1.5pt]{};

            \draw[->,gray,thick] (-0.25,0) -- (0.25,0);

            \draw[thin,gray] (1,0.1) -- (1,0.5);
            \draw[thin,gray] (1.7,-0.1) -- (1.7,-0.5);
            \draw[thin,Maroon] (2.4,0.1) -- (2.4,0.85);
            \draw[thin,gray] (3.1,-0.1) -- (3.1,-0.5);

            \draw[thin] (1,0.1) -- (1.7,-0.1);
            \draw[thin] (1.7,-0.1) -- (2.4,0.1);
            \draw[thin] (2.4,0.1) -- (3.1,-0.1);
        
		\node at (1,0.1) [circle,fill,inner sep=2pt]{};
            \node at (1.7,-0.1) [circle,fill,inner sep=2pt]{};
            \node at (2.4,0.1) [circle,fill,inner sep=2pt]{};
            \node at (3.1,-0.1) [circle,fill,inner sep=2pt]{};

            \node at (1,0.5) [circle,fill,gray,inner sep=1.5pt]{};
            \node at (1.7,-0.5) [circle,fill,gray,inner sep=1.5pt]{};
            \node at (2.4,0.85) [circle,fill,Maroon,inner sep=1.5pt]{};
            \node at (3.1,-0.5) [circle,fill,gray,inner sep=1.5pt]{};

            \node at (2.05,-0.95) {all other nuclei fixed};
            \draw[->,semithick,Maroon] (2.4,1.) -- (2.4,1.3);
	\end{tikzpicture}
    \end{figure}
    
    All nuclei, except the Tritium of interest, are kept fixed at their original equilibrium positions. This is equivalent to consider the motion of the Tritium of interest as faster than that of all other nuclei.
\end{itemize}
\begin{table*}
\begin{tabular}{c|c|ccccccc|c|c}
Coverage & Configuration & $M_0$   & $M_{f}$   & $z_0$ &$z_{b}$ & $E_{d}$ & $E_{a}$ & $E_{b}$ & Prescription & T motion   \\[-0.4em]
(\%)&  &($\mu_{\rm B}$)&($\mu_{\rm B}$)& (\AA) &(\AA) &(eV) &(eV)  &(eV)  & for relaxation & method \\
\hline\hline
100 & chair & 0   & 2 &  1.10 &  -- & 4.35& 0.00 & $-4.35$ & one C fixed & static \\
100 & chair & 0   & 2 &  1.10 & 3.65 & 4.65& 0.30 & $-4.35$ & one C fixed & steered \\
100 & chair & 2   & 2 &  1.05 & 2.61 & 1.50& 1.00 & $-0.50$ & one C fixed & steered \\ 
100 & chair & 0   & 2 &  1.10 & 2.97 & 5.27& 0.42 & $-4.85$ & all C fixed & steered   \\
100 & chair & 2   & 2 &  1.10 & 1.80 & 1.69& 1.02 & $-0.67$ & all C fixed & steered \\
100 & chair & 0   & 2 &  1.10 & -- & 4.62& 0& $-4.62$ & all fixed & static \\
50  & chair & 16  & 16&  1.15 &1.65& 0.71& 0.37 & $-0.34$ & one C fixed & steered \\
6.2 & $D=5.8~\text{\AA}$ trans&0&2&1.11&2.01&1.64&0.33& $-1.31$ & one C fixed & steered \\
6.2 & $D=4.4~\text{\AA}$ trans& 0& 2 & 1.13&       &       &       &$-0.39$  & all relaxed & static  \\
6.2 & $D=4.4~\text{\AA}$ trans& 2& 2 & 1.13  &   &       &            &$-0.60$  & all relaxed & static \\
6.2 & $D=4.4~\text{\AA}$ cis  & 0& 2 & 1.12  &       &       &       &$-0.37$  &  all relaxed& static  \\
6.2 & $D=4.4~\text{\AA}$ cis  & 2& 2 &1.13   &       &       &       &$-0.54$  &  all relaxed & static   \\
6.2 & \scriptsize{$D=1.4~\text{\AA}$ ortho-trans}&  0& 2 & 1.12 &  3.1  &2.55 & 0.19 & $-2.36$ & one C fixed & steered \\
6.2 & \scriptsize{$D=2.5~\text{\AA}$ meta-trans} &  2& 2 & 1.13 &  2.5  &0.89 & 0.29 & $-0.60$ & one C fixed & steered \\
6.2 & \scriptsize{$D=2.9~\text{\AA}$ para-trans}&  0& 2 & 1.12 &  2.0  &1.93 & 0.36 & $-1.57$ & one C fixed & steered \\
6.2 & \scriptsize{$D=1.4~\text{\AA}$ ortho-cis} &  0& 2 & 1.12 & -- & 2.50 & 0.7 & $-1.79$ & prevent ${\rm T}_2$ & steered \\
6.2 & \scriptsize{$D=2.5$\AA\ meta-cis}  & 2 & 2 & 1.13 &  2.5 &0.92 & 0.27  & $-0.65$ & one C fixed & steered \\
6.2 & \scriptsize{$D=2.9$\AA\ para-cis}  &  0& 2 &1.12& 2.23  &2.07 & 0.32 & $-1.75$ & one C fixed & steered \\
3.1 & $D=10.2$~\AA, cis  & 0& 1 &1.12&1.75& 1.18&  0.48 & $-0.70$ & one C fixed & steered \\
3.1 & $D=10.2$~\AA, cis  & 1& 1 &1.12&1.75& 1.23&  0.48 & $-0.75$ & one C fixed & steered \\
\end{tabular}
\hspace{1.5em}
\begin{tabular}{c}
\includegraphics[width=0.2\textwidth]{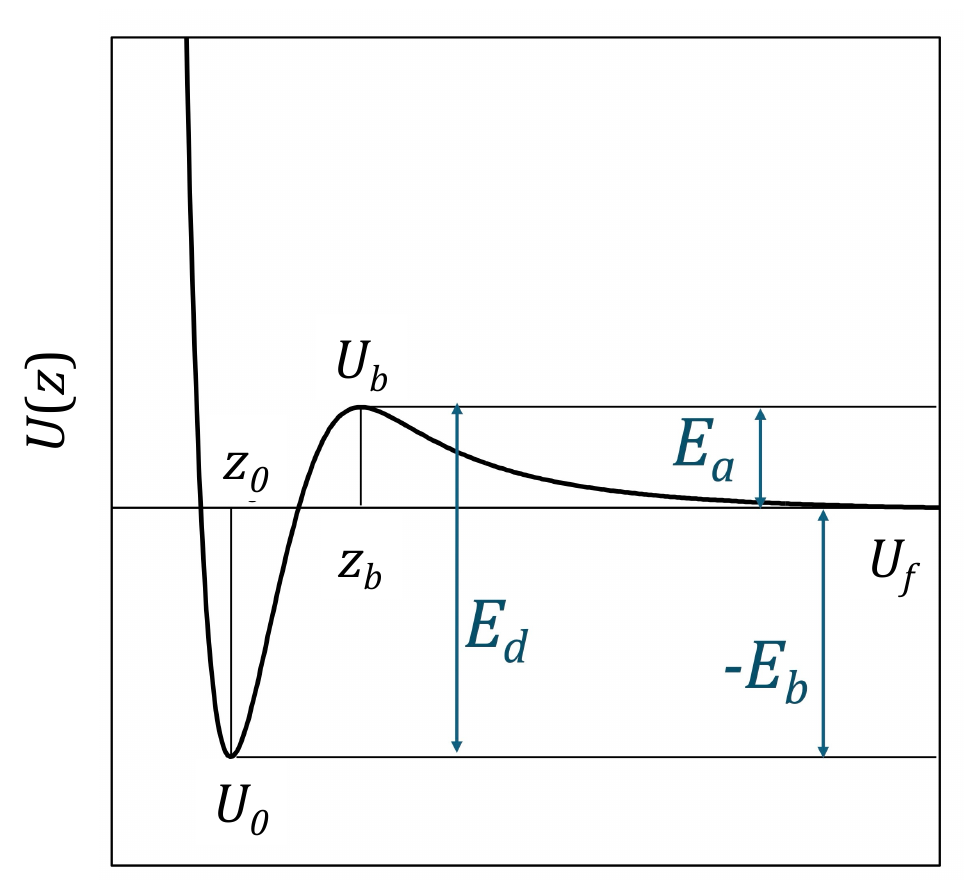}\\
\includegraphics[width=0.2\textwidth]{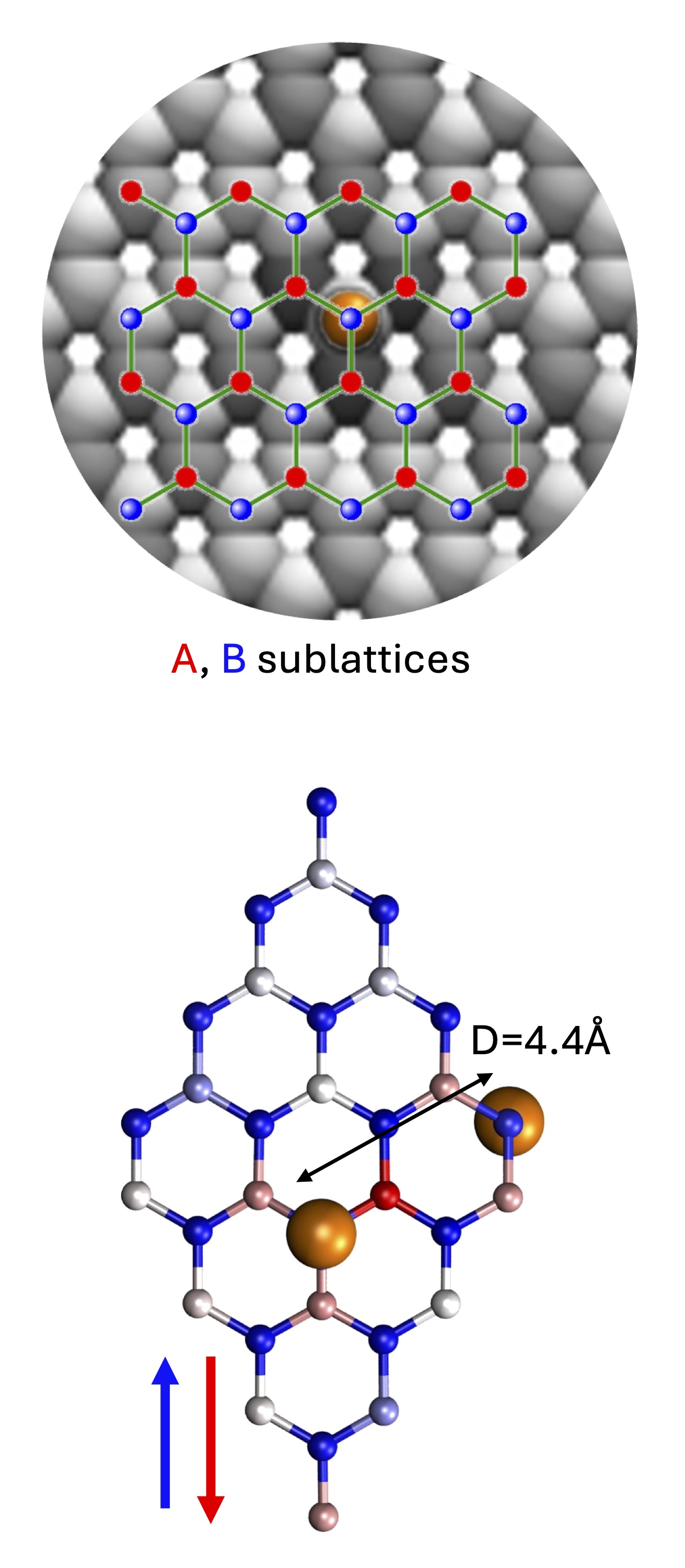}\\
\end{tabular}
\caption{{\bf Table:} Summary of energies, magnetizations and some structural parameters evaluated within the BO-DFT scheme for the initial Tritium orthogonal potential, under different combinations of loading, spatial configuration and magnetization. The naming of the parameters refers to the first figure on the top right (see also the main text). The coverage in the first column is the T:C percentage. The total magnetization of the bound state, $M_0$, and of the unbound state, $M_{f}$, in the supercell (always 4$\times$4 repeatition of the unit cell) are expressed in units of the Bohr magneton, $\mu_{\rm B}=e\hbar/2m_e$ (in these units, the intrinsic electron magnetic moment is $\simeq 1$). In the last two columns we report the prescription used for the other nuclei, and the method employed to implement their relaxation, as explained in Sec.~\ref{sec:rest}. Specifically the last column reports the method used to move the Tritium. With the ``static'' method, the Tritium is fixed at a given position (which usually takes 30 values, evenly spaced) along the path. In each position, the electronic calculation is performed re-initializing the system. With the ``steered'' method, instead, the Tritium is pulled or pushed with an external force, while at the same time introducing a damping term in Newton's equations, to makes sure it moves very slowly and continuously along the path. In spite of the fact that many more points are evaluated, this method can be faster in some circumstances because at each step the electronic structure changes only by little, and can thus be use as the starting point for the subsequent step of the self-consistent calculation. 
In the cases of 6.2\% loading, obtained including 2 atoms in the supercell, $D$ is the distance of their location on the graphene surface. For $D=4.4~\text{\AA}$, only the final structures were evaluated, and therefore the location and energy of the barriers are not available. In the case of 3.1\% loading a single atom is loaded on the structure, and therefore the distance with other atoms correspond to the edge of the supercell, i.e. $\sim \! 10.1 \text{ \AA}$. In the orto-cis case we manually added a restriction on the T-C  distance of the second Tritium atom, in order to prevent the dissociation to a full ${\rm T}_2$ molecule, which is competitive with single Tritium dissociation in this case. For this case, the location of the barrier is not well defined, and therefore not reported. {\bf Upper figure:} schematic representation of the general form of the potentials with the definition of the various quantities reported in the table and in the main text. {\bf Middle figure:} representation of the graphene lattice with a Tritium bound to it. The gray shaded regions correspond to the level of magnetization around the bound Tritium. The two sub-lattices, A and B, are also reported. {\bf Lower figure:} graphene lattice with two Tritium atoms bound to the same sub-lattice. The Carbon sites are colored in blue or red depending on whether they present an up or down magnetization.
}\label{tab:energies}
\end{table*}
For the first two options, we further considered two different ways of moving the the coordinate $\bm X_{\rm T,1}$: either we fix it at increasing values to emulate the motion of Tritium/Helium out of the structure, and relax the other atoms as prescribed, or pull/push it slowly with a small external force, letting the others relax as prescribed. The different option return slightly different curves as reported in {\color{Maroon}Fig.\ 3} in the main text.
\subsection{The Tritium orthogonal potential: barriers, desorption energies and other details}
\noindent The potentials were evaluated combining the prescription given above, different coverage levels and magnetization, as per full list reported in Table~\ref{tab:energies}. The table reports the potential depths, $E_b$, desorption and adsorption barriers, $E_d$ and $E_a$, as defined in the main text Section~{\color{Maroon} III.B}, and sketched in the plot aside the table. We also show schemes of the spin distributions in specific cases, as specified in the caption. 
\subsection{Comparison of boat vs chair graphane}\label{app:boat}
\noindent Table~\ref{tab:boat} reports the structural parameters and energies of graphene in boat conformation (fully and single side half loaded, as depicted under the table) compared to the chair conformation. The fully loaded conformation is more stable in non-magnetic state, and the chair conformation is approximately 0.1~eV per (CH) unit more stable with respect to the boat one. In the 50\% loading (one sided), the magnetized state is more stable in the chair conformation, which is 0.11~eV more stable than the non-magnetic one, although the non-magnetic boat state appears further stabilized by 0.6~eV. 
\begin{table}[t]
\begin{tabular}{ccccccc}
Cov. & Config. & $M$/C & $d_{\text{C-H}}$ & $d_\text{C-C}$ & $\Delta E$&$\Delta E$/C \\  
(\%)&     & ($\mu_{\rm B}$)            &(\AA)      &(\AA)      & (eV)      &(eV)\\
\hline
\hline
100 & chair & 0     & 1.11      & 1.53      & 0 (ref) & 0 (ref)\\
100 & boat      & 0     & 1.10      &1.53, 1.56 & 3.2     & 0.10 \\
\hline
50  & chair & 0     & 1.16      &1.50       & 0 (ref) & 0 (ref)\\
50  & chair & 1      & 1.15      &1.50       &$-3.7$     & $-0.11$ \\ 
50  & boat      & 0     & 1.13      &1.51, 1.36 &$-5.5$     & $-0.17$
\end{tabular}
\begin{tabular}{c}
\includegraphics[width=3.3cm]{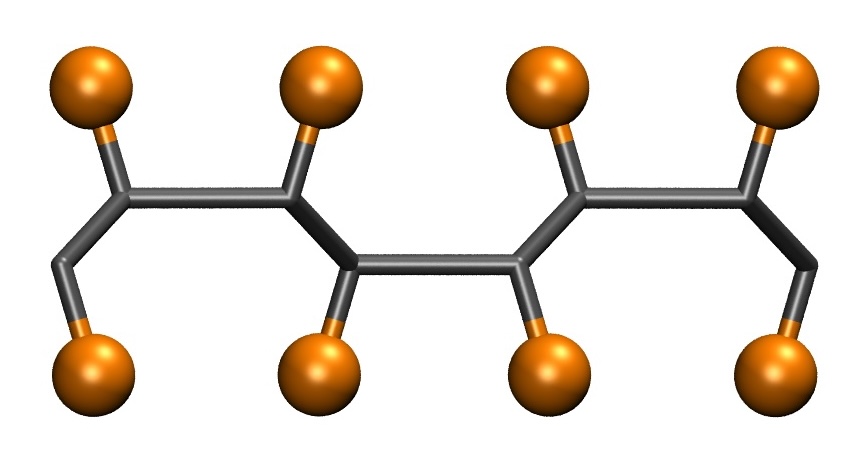}\\
{\scriptsize boat 100\%, side view}\\
\includegraphics[width=3.3cm]{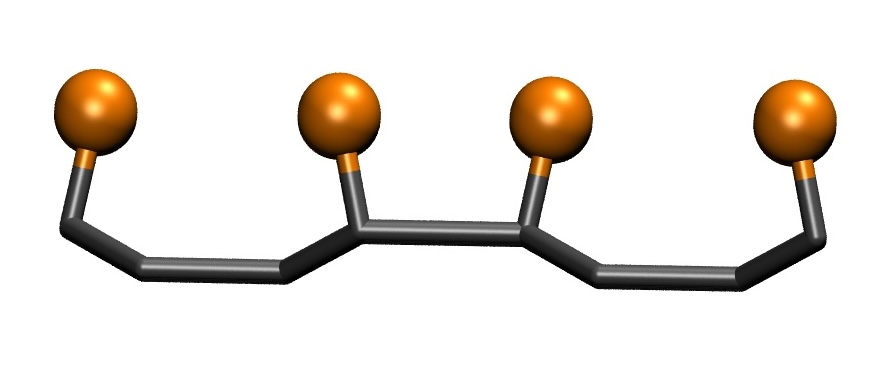}\\
{\scriptsize boat 50\%, side view}
\end{tabular}
\begin{tabular}{c}
\includegraphics[width=4.6cm]{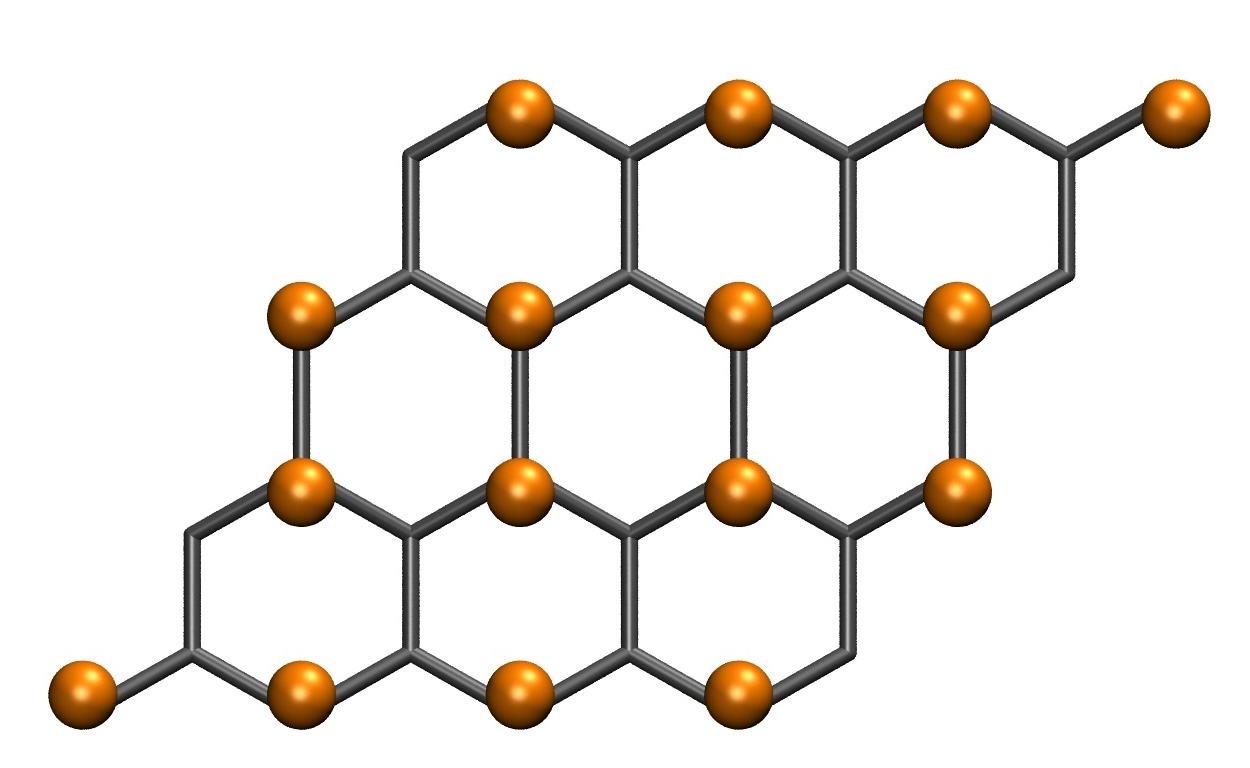}\\
{\scriptsize boat 100\%, top view}
\end{tabular}
\caption{Energies, magnetization and geometric parameters of the relaxed structures of fully and one sided loaded graphene in chair and boat conformation (illustrated under the table). The relaxed parameters of the $4\times 4$ supercell of the chair conformation  are  $10.125\times10.125\times40$~\AA$^3$ (hexagonal symmetry), i.e. $2.53~\text{\AA}$ per unit cell, at 100\% loading, with no appreciable variation at 50\% loading. In the case of fully loaded boat conformation, the symmetry deviates slightly from hexagonal, and the parameters are $10.74\times9.96 \text{ \AA}^2$ (with an angle of $59.62^\circ$ instead of $60^\circ$). Tritium atoms are in orange, C-C bonds are in gray.}
\label{tab:boat}
\end{table}
\subsection{Electronic relaxation and energetics of Helium release}
\begin{figure}[t]
\scriptsize{Electronic relaxation, top view}\\
\includegraphics[width=0.6\linewidth]{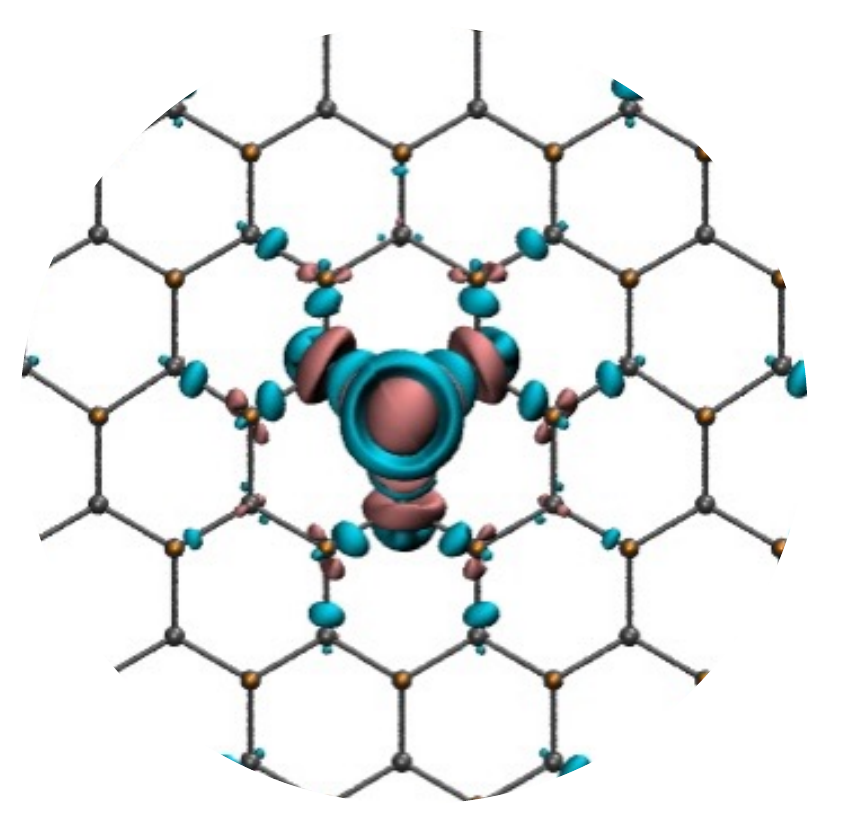}\\
\begin{tabular}{cccccc}
        &System & Charge& $\Delta E_{\rm relax}$  & $d_\text{C-He}$ & $M$ \\
        &     &    & (eV)                &(\AA)          & ($\mu_{\rm B}$)\\
\hline\hline
\includegraphics[width=0.2\linewidth]{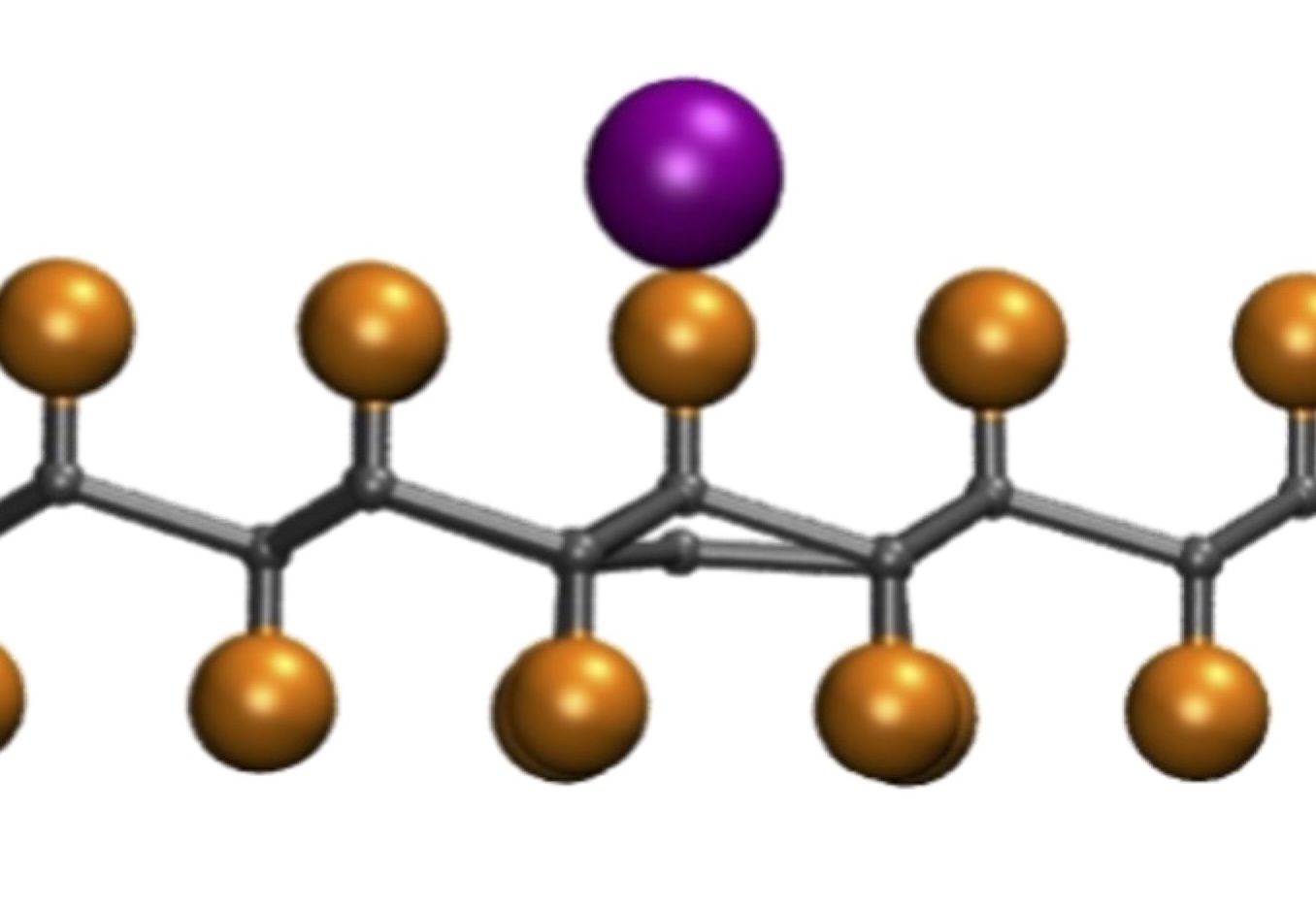}&1 He & +1  & $-2.72$              & 2.74  & 0  \\
\includegraphics[width=0.2\linewidth]{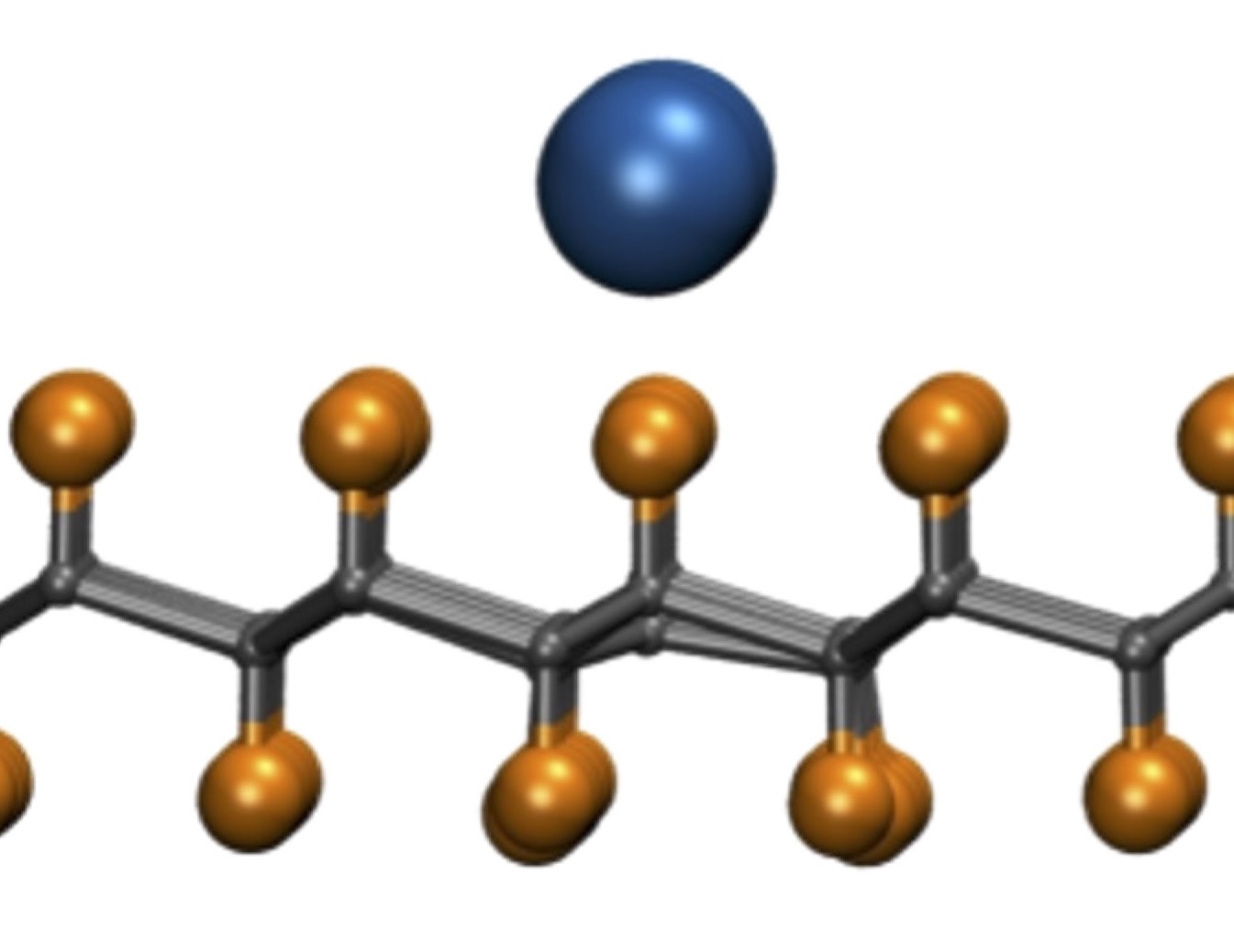} &1 He & 0  & $-5.18$              & 3.36  & 1  \\
\includegraphics[width=0.2\linewidth]{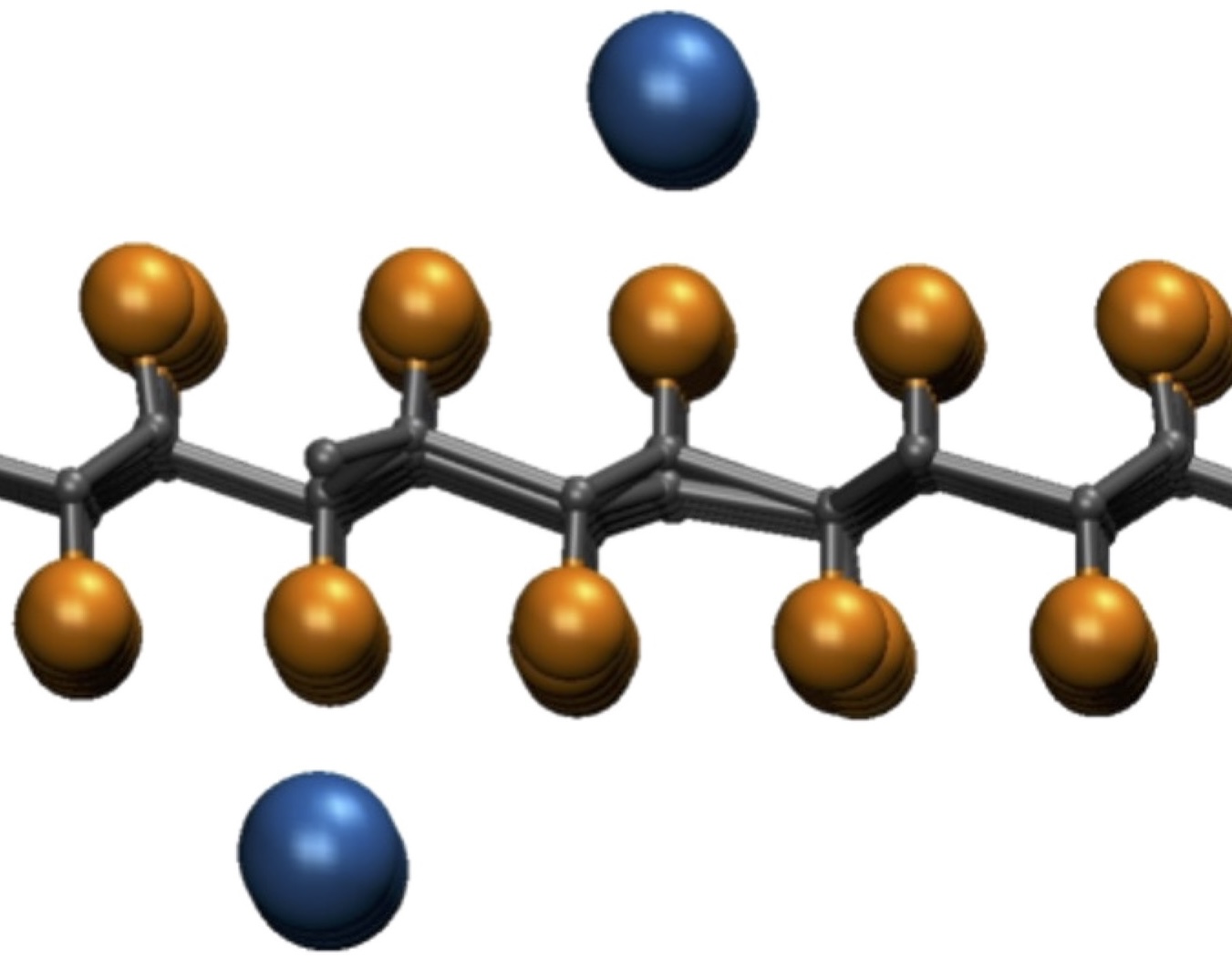}&2 He & 0   & $-6.42\times2$     & 3.41  & 2 \\          
\end{tabular}
\caption{{\bf Top figure:} A representation of the iso-surfaces of the electronic structure displacement upon electronic relaxation after the ${\rm T}\to {\rm He}$ transition. The difference of electronic density before and after relaxation $\Delta n(\bm x)$ is evaluated and represented as iso-density surfaces. Light blue surfaces enclose areas where $\Delta n$ is negative, i.e. where the electron charge is depleted, while pink surfaces enclose areas where the electron density increases. {\bf Table:} Locally relaxed structures in three different cases (isolated Helium in electrically isolated or neutralized system, and two released Helium in the the neutralized system), with the corresponding structural relaxation energies, structural parameters and magnetization. }\label{tab:localopt}
\end{figure}

\noindent After the ${\rm T}\to{\rm He}$ transition, two kinds of relaxations occur: first a fast electronic relaxation, and second a slower structural relaxation. The electronic relaxation is represented in the top part of Fig.~\ref{tab:localopt}, showing that, as the electronic structure goes back to the ground state, electrons tend move towards the Helium from the neighboring sites, to compensate its extra nuclear charge. The relaxation energy can be approximately evaluated as the energy difference between the sudden approximation and the adiabatic one with the Helium in $\bm z_0$, and turn out to be $\sim \!50$~eV.
%
%

After the electronic relaxation, the Helium is repelled by the substrate, as shown from the Molecular Dynamics (MD) simulations in the main text (Section {\color{Maroon} VI}), with an energy gain of the system of 2-4~eV, depending on the charge state. However, the quantification of the relaxation energy is better accessible to a geometry optimization simulation, where the system is fully damped to a local minimum of the energy. This is a state in which the Helium stops in a very shallow potential well defined by the van der Waals attraction between itself and  the substrate. This is not observed in the MD simulation because the Helium is released with high kinetic energy, thus easily overcoming the weak attraction. The positions corresponding to these shallow potential wells are illustrated in Fig.~\ref{tab:localopt} for the two charge states of a single detached Helium, and for a couple of nearby Helium, released on opposite sides of the sheet. The energy gain upon detachment, and the location of the Helium in this van der Waals well are larger in the neutral state, and further increase in the case of nearby detachment, showing a cooperative effect.

It is also interesting to look at the magnetic state of the system. After the detachment of Helium, the +1 charged system is left with an even number of electrons and stays in a null magnetic state. When it is neutral, conversely, it is left with an odd number of electrons, with the unpaired one located in the dangling bond. The magnetization is thus $M=1\mu_{\rm B}$. We also analyzed the case in which two Helium atoms are detached, and the system is kept neutral. In this case the sheet is overall neutral, but has two extra electrons with respect to the initial configuration, which appear to be in triplet state, with $M=2\mu_{\rm B}$. 
%
\begin{figure}[t]
\includegraphics[width=0.9\linewidth]{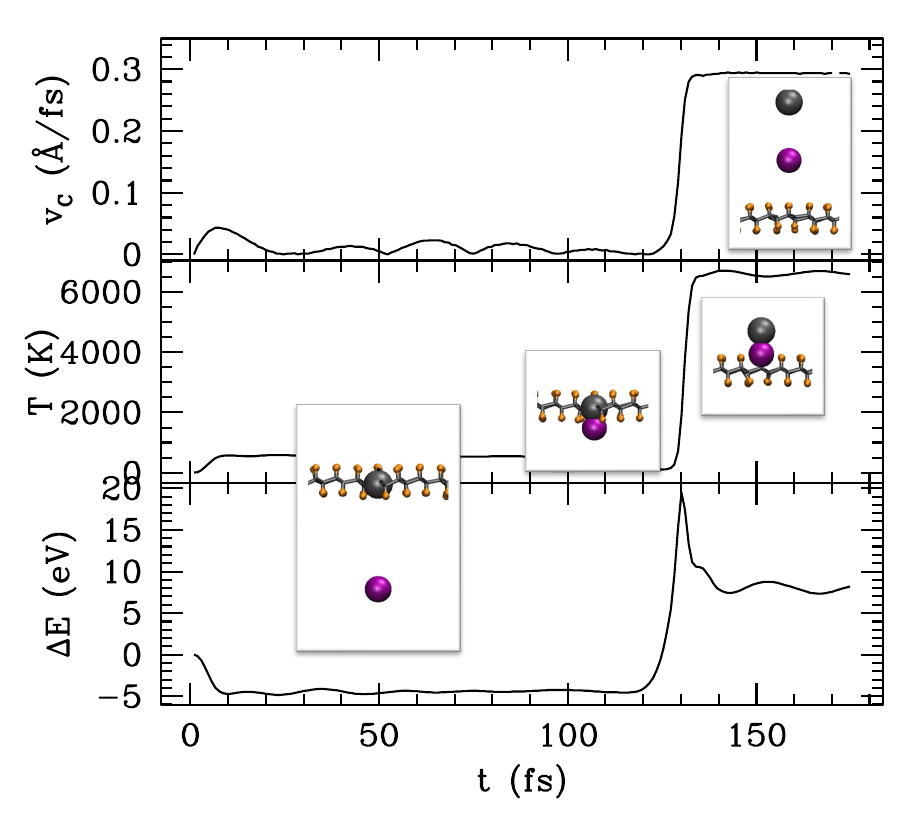}
\caption{Simulation of the frontal collision of the Helium on a vacant Carbon site. The Helium is accelerated to an energy of $\sim \! \text{eV}$ as the result of expulsion from another sheet, before hitting the Carbon site. From bottom to top energy variation, temperature and velocity of the ejected Carbon. The Helium is represented as a purple sphere, while the Carbon is the dark gray sphere. }\label{fig:hit}
\end{figure}

Finally, Fig.~\ref{fig:hit} reports a specific instance, created ad hoc in order to show under which conditions the Helium, after expulsion, can create a defect in another sheet. The Helium is expelled with energy of about 3~eV, and then hits frontally an already vacant Carbon site on another sheet. We choose the vacant Carbon site such that it is surrounded by T-loaded sites, as the bonds between a vacant site and loaded sites are weaker. In this instance, the collision between the Helium and the Carbon can cause the extraction of the latter, creating a defect. A movie of the event is reported as supporting material, while selected snapshots are in the plot. As seen from the plots, the energy of the system increases by several eVs as the vacancy forms and the extracted Carbon atom acquires velocity.
\subsection{Implementation of the (Semi-)Sudden Approximation }
\noindent The dependencies of the ground state energy on the nuclear coordinates, $\{\bm X\}$, and on the electron density, $n(\bm x)$, are described by Eq.~({\color{Maroon} 7}) of the main text, where $E_0$ is expressed as the sum of 
$E^{\rm int}[n]$, including both the kinetic energy and electron-electron interactions, and of $E^{e{\rm N}}[n,\{\bm X\}]$ and $E^{\rm NN}[\{\bm X\}]$, which are instead the electron-nucleus and nucleus-nucleus interactions. 
In the standard DFT scheme, $n$ is determined self-consistently for any given set of nuclear coordinates, i.e. $n\equiv n_{\{\bm X_{\rm T/He}, \bm X \}}(\bm{x})$, where $\bm X_{\rm T/He}$ is the coordinate of the Tritium/Helium under consideration, while $\{\bm X\}$ are the coordinates of all other nuclei. Indeed, this is how we determined the potentials, $U$, for the Tritium right before the decay, and for the Helium well after relaxation, when the BO approximation is valid: we evaluated the PES by only varying $\bm X_{\rm T/He}$.

In our implementation of the sudden approximation, instead, $n(\bm x)$ is kept frozen to the configuration it had before the decay, and no self-consistent cycle is performed. Consequently $E^{\rm int}$ is unchanged, while the $E^{e{\rm N}}$ and $E^{\rm NN}$ must be recalculated to account for the transition ${\rm T}\to{\rm He}$, with the consequent change of nuclear charge. $E^{\rm NN}$ is a simple sum over the nuclei in the cell, i.e.,
\begin{align}
    \begin{split}
        E^{\rm NN}(\bm X_{\rm He},\{\bm X\})={}& \frac{1}{2}\sum_{IJ}\frac{Q_IQ_J}{|\bm{X}_I -\bm{X}_J|} \\
        & + E^{\rm vdW}(\bm X_{\rm He},\{\bm X\}) \,,
    \end{split}
\end{align}
which also includes the van der Waals dispersion corrections. The periodicity in the evaluation of the long range Coulomb interaction is accounted for with the Ewald summation method, as implemented in the Quantum Espresso (QE) code~\cite{giannozz_01}. 

On the other hand, the evaluation of  $E^{e\rm N}$ is not a standard output of a QE calculation, and required some post-processing. In particular, we must compute,
\begin{align}
    \begin{split} \label{eq:EeN}
        E^{eN}[n,\{\bm X_{\rm He},\bm X\}]={}& \int d\bm x \,  n({\bm x}) \sum_I \left[-\frac{Q_I}{|\bm{x} -\bm{X}_I|}\right]\\
        \equiv {}&\int d\bm x \,  n({\bm x}) V_{\bm X_{\rm He},\{\bm X\}}(\bm x)  \,,
    \end{split}
\end{align}
where $V_{\bm X_{\rm He},\{\bm X\}}(\bm x)$ is returned by QE at any given configuration of the Helium along the path. As far as $n(\bm r)$ is concerned, we consider two possible schemes (see also main text): 
(i) in the very {\it sudden approximation}, we take 
$n(\bm x) \equiv n_{\bm X_{\rm T} = (0,0,z_0)}(\bm x)$, i.e. the density evaluate just before decay, with the Tritium in its equilibrium position, corresponding to the minimum of the binding well; (ii) in the {\it semi-sudden} approximation, we instead take $n(\bm x) \equiv n_{\bm X_{\rm T} = \bm X_{\rm He}}(\bm x)$, which is again the density evaluated just before the decay, but for a Tritium position corresponding to the current Helium coordinate.

All the relevant quantities --- i.e., $V_{\bm X_{\rm He},\{\bm X\}}(\bm x)$, $n_{\bm X_{\rm T} = (0,0,z_0)}(\bm x)$ and $n_{\bm X_{\rm T} = \bm X_{\rm He}}(\bm x)$ --- are functions of a spatial coordinate, $\bm x$, and can be extracted from QE calculations on space grids. Consequently, in both instances described above, the corresponding integral in Eq.~\eqref{eq:EeN} is performed summing on each grid point, using the Python package Cube~Toolz~\cite{cubetoolz}.
\subsection{Alignment of the sudden, semi-sudden and adiabatic potentials}
\noindent The relative offset for the Tritium and Helium potentials cannot be directly inferred from DFT calculations, because the pseudopotentials used for different kinds of atoms have generally different references. However, for large enough distances ($\gtrsim 10\!-\!15$~\AA), the sudden potentials essentially match a pure Coulomb interaction of He$^{++}$ with graphene (which bears charge $-1$). At intermediate distances --- about $\!3-\!5~\text{\AA}$ ---, where our data lie, the curve is better represented by the interaction of two displaced pointlike charges: a $-2$ charge representing the doublet deriving by the (frozen) T--C bond, and the charge of the nearest Carbon nucleus, that is +1 due to the screening of its remaining 5 electrons (all the other nuclei of the structure are well screened by the electronic structure and, therefore, essentially neutral). Consequently, the potential in this region is well represented by the formula,
\begin{align}
U^{\rm Coul.}_{\rm He}=-e^2\frac{2\times2}{|\bm{x}-\bm{x}_{\rm bond}|}+e^2\frac{2}{|\bm{x}-\bm{x}_{\rm C}|} \,,
\end{align}
with the displacement between the Carbon nucleus and the bond position given by approximately of 0.7~\AA. The excellent quality of this representation is shown by its superposition (blue dotted line) onto the sudden approximation potential (solid blue line in Fig.~\ref{fig:si4}), in the region of interest. At shorter distances ($\lesssim 2$~\AA), the non-pointlike nature of the electronic distribution of the bond doublet becomes relevant and $U_{\rm He}^{\rm Coul.}$ is no longer a good approximation to the true potential. We therefore used the Coulomb potential to establish the reference energy of the sudden approximation potential which, by definition, vanishes when the He$^{++}$ ion is infinitely far away from the graphene. We then rigidly shift the semi-sudden (magenta) and adiabatic (blue) potentials accordingly. 
\begin{table}
    \centering
    \includegraphics[width=0.95\linewidth]{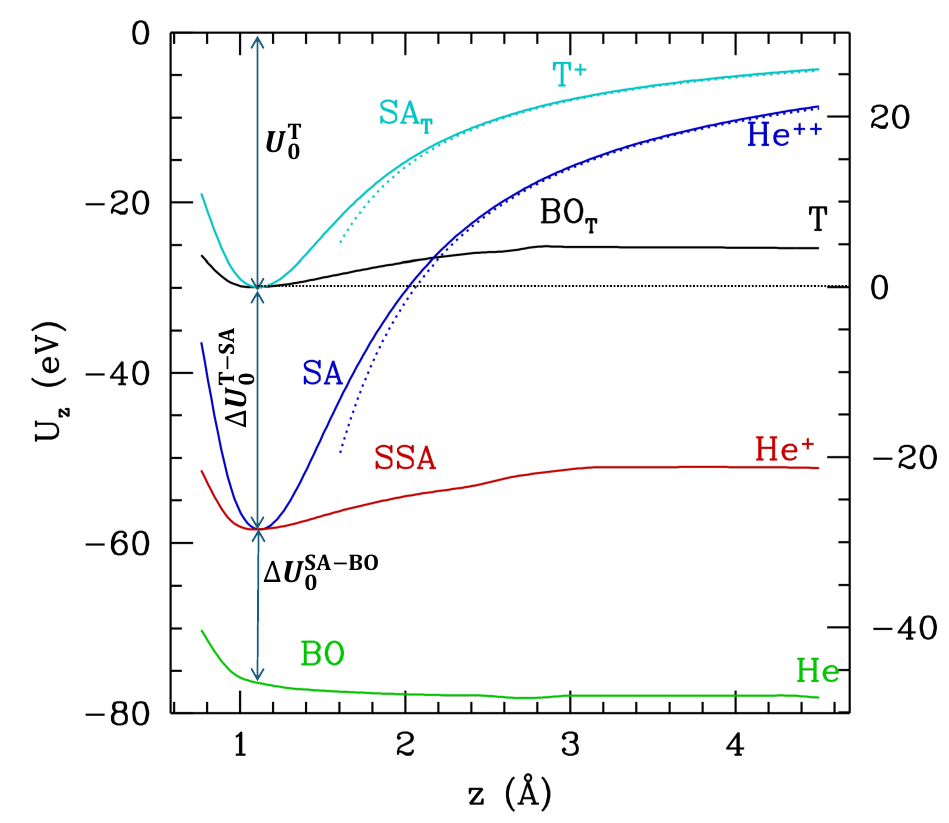}
    \begin{tabular}{c|c|c|c}
        Potential & $U_0$ [eV] & $U_\infty$ [eV] & $E_b\equiv U_\infty - U_0$ [eV] \\\hline\hline
        Adiabatic T & $-29.9$ & $-25.3$ & $4.6$ \\
        Sudden T & $-29.9$ & $0$ & $29.9$ \\
        Sudden He & $-58.5$ & $0$ & $58.5$ \\
        Semi-sudden He & $-58.5$ & $-51.1$ & $7.4$ \\
        Adiabatic He & $-76.1$ & $-78.3$ & $-2.2$ 
    \end{tabular}
    \caption{Adiabatic (BO), sudden (SA) and semi-sudden (SSA) potentials for Helium (green, blue and red), as well as adiabatic and sudden potentials for Tritium (black and cyan). The potentials are aligned using the electrostatic zero, obtained with He$^{++}$ and T$^+$ (blue and cyan dotted lines) infinitely far way, as explained in the text. For convenience, also the original energy scale of the Tritium potential with its reference to the bound state, is reported on the vertical axis on the right. A table with numerical values of specific energy levels and their differences is reported under the plot. 
    }\label{fig:si4}
\end{table}

To apply the same criterion to the Tritium potential, we computed, as an intermediate step, the sudden potential for Tritium (obtained moving Tritium away from the structure while keeping frozen all other atoms and electrons). This one is also assumed to vanish when the Tritium is infinitely far away from the graphene, just like it does for the He$^{++}$ ion. Consistently, we consider the displaced Coulomb form,
\begin{align}
    U^{\rm Coul}_{\rm T}=-e^2\frac{2}{|\bm{x}-\bm{x}_{\rm bond}|}+e^2\frac{1}{|\bm{x}-\bm{x}_{\rm C}|} \,.
\end{align}
Similarly to what happens for the case of the Helium sudden potential, the expression above provides a good description of the actual sudden Tritium potential in the range $3\!-\!5~\text{ \AA}$ (cyan solid and dotted curves), while it fails for smaller distances. As we did for Helium, the Tritium adiabatic potential was shifted rigidly. This finally sets the same reference alignment for all potentials, and fixes the energy difference between the bound state of Tritium and the ``sudden'' state of Helium (just after the decay) to be about $\Delta U_0^{\rm T-SA} \equiv U_0^{\rm T}-U_0^{\rm SA}\simeq 29$~eV (difference between the bottom of cyan-black and red-blue curves), while about $\Delta U_0^{\rm SA-BO}\equiv U_0^{\rm SA}-U_0^{\rm BO}\simeq 18$~eV are gained upon electronic relaxation (vertical distance to the green curve). 
\section{Solution to the Helium nuclear Schr\"odinger equation and computation of the event rate} \label{app:calc}
\begin{figure*}
    \centering
    \includegraphics[width=\linewidth]{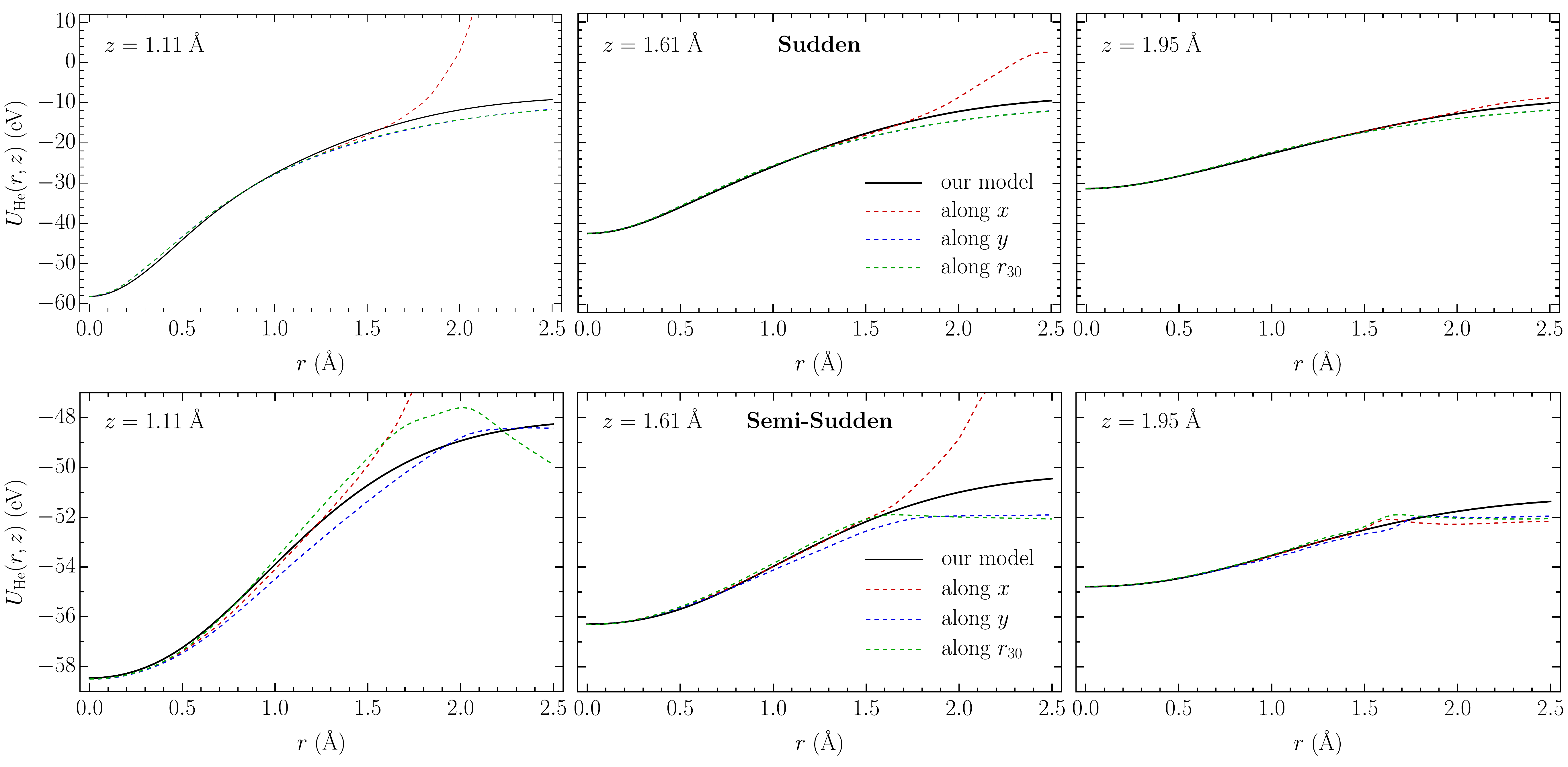}
    \caption{Helium potentials along the three in-plane directions explored in this work, for different values of $z$. Dashed lines show the results of the DFT calculation, for both the sudden ({\bf upper panels}), and semi-sudden ({\bf lower panels}) schemes. We also show the semi-analytical model of Eq.~\eqref{eq:potential_pheno}. The Helium potential obtained within the adiabatic scheme does not admit bound states, and its repulsive profile is shown in Fig.~{\color{Maroon} 4} of the main text. A gray dotted line has also been added, to guide the eye towards the asymptotic zero of the potential, as $r \rightarrow \infty$. Solutions to the Schr\"{o}dinger equation whose eigenvalues are negative constitute the discrete spectrum, whereas solutions with positive eigenvalues belong to the continuous spectrum. Finally, we stress that these curves have been offset in order for them to indeed vanish at infinity. This is a different choice that what done in Fig.~{\color{Maroon} 4} of the main text.}
    \label{fig:potHemodel}
\end{figure*}

\begin{table*}
\begin{tabular}{c|cccc|cccc|cccc}
\multicolumn{1}{c|}{} & \multicolumn{4}{c|}{{\bf Sudden He}} & \multicolumn{4}{c|}{{\bf Semi-sudden He}} & \multicolumn{4}{c}{{\bf Adiabatic T}} \\
$z$ \!\big(\AA\big) & $k$ \!\big(eV\AA${^{-2}}$\big) & $S$ \! \big(eV\big)   & $U_z$ \!\big(eV\big)   & $W$ \!\big(\AA$^{-2}$\big) & $k$ \!\big(eV\AA${^{-2}}$\big) & $S$ \! \big(eV\big)   & $U_z$ \!\big(eV\big)   & $W$ \!\big(\AA$^{-2}$\big) & $k$ \!\big(eV\AA$^{-2}$\big) & $S$ \! \big(eV\big)   & $U_z$ \!\big(eV\big)   & $W$ \!\big(\AA$^{-2}$\big) \\
\hline\hline
1.11 &  152.37 & $-8.5$  & $-58.5$ &  0.55       &  8.94 & $-48.05$ & $-58.5$    & 0.63     & 5.44  & 6.33 & 0    & 0.63\\
1.61 &  65.80  & $-8.5$  & $-42.78$ &  0.49   &  4.45 & $-50.21$  & $-56.3$ & 0.52     & 3.33  & 5.88 & 1.56 & 0.52  \\
1.95 &  29.67  & $-8.5$  & $-31.65$ &  0.38    &  2.47 & $-51.15$  & $-54.79$ & 0.45    & 2.84  & 5.70 & 2.77 & 0.45 
\end{tabular}
\caption{Fitted values of the functions appearing in the definition of the potential $U(r,z)$ in Eq.~\eqref{eq:potential_pheno} for three distinct values of $z$ probed with our DFT simulations, both for Tritium and Helium. Different parameters are obtained depending on both the considered nucleus and the adopted scheme (sudden, semi-sudden, and adiabatic), leading to the plots shown in Fig.~\ref{fig:potHemodel} for the case of Helium; the parameters found for Tritium, instead, allow to reproduce the curves plotted in Fig.~{\color{Maroon}3} in the main text for the three considered values of $z$. The constant $P$ has been set to $2.8 \text{ \AA}^{-2}$, $3.6 \text{ \AA}^{-2}$ and $3.6 \text{ \AA}^{-2}$ for the three cases, respectively.} 
\label{tab:pars}
\end{table*}
\noindent In the first part of this section we show how to convert the raw DFT data describing the $U_{\mathrm{T}}$ and $U_{\mathrm{He}}$ profiles (shown respectively in Fig.~{\color{Maroon}3} in the main text and in Fig.~\ref{fig:potHemodel}) into a function, $U(\bm{x})$, that can then be used to numerically solve the Schr\"{o}dinger equation for the nuclei, as explained in Section~{\color{Maroon} V} of the main text.
As argued in Section~{\color{Maroon} III.C} in the main text, both $U_{\mathrm{T}}$ and $U_{\mathrm{He}}$ exhibit an approximate cylindrical symmetry, for regions close to their minimum, so that $U(\bm{x}) \simeq U(r,z)$. We then use the following model to interpolate the DFT data,
\begin{align} \label{eq:potential_pheno}
\begin{split}
     U(r,z)= S(z)+ \Biggl\{ U_z(z)-S(z)
     + \left[ \frac{1-e^{-Pr^2}}{P} \right] \\
     \times \biggl[ W(z)\bigl[U_z(z)-S(z)\bigr] 
     +\frac{k(z)}{2} \biggr] 
       \Biggr\} \, e^{-W(z)r^2} \, .
\end{split}
\end{align}

Here, $S(z)$ is the asymptotic value of the potential at $r \to \infty$, where the dependence on $z$ is introduced to account for the $z$-dependent values of the repulsive barrier which, in the region of interest for our calculation, is only partially probed. 
On the other hand, for small $r$ one may expand the expression in Eq.~\eqref{eq:potential_pheno} to obtain $U(r\rightarrow0,z) = U_z(z) + \frac{1}{2}k(z)r^2 +\mathcal{O}(r^4)$, where $U_z(z) \equiv U(r=0,z)$, and $k(z)$ provides the first harmonic corrections around the minimum of the potential, with a spring constant whose value depends on the distance from the layer. Finally, the auxiliary function $W(z)$ and the constant $P$ have been tuned to properly connect the small and large $r$ regimes. These functions have been fitted for the three values of $z$ at which $U(r,z)$ has been explored with DFT, providing the parameters reported in Table~\ref{tab:pars}, for the sudden and semi-sudden schemes for the Helium, and for the adiabatic scheme for the Tritium. These fits, in turn, allow to extrapolate 
the function $U(r,z)$ in the whole volume of interest, and thus solve
the Schr\"odinger equation,
in order to obtain the wave function $\psi_{i,\mathrm{T}}$ describing the initial state of the system, and the wave function $\psi_{f,\mathrm{He}}$ describing its final state. 

The remainder of this section is devoted to the computation of two distinct contributions to the $\beta$-decay event rate: one due to final Helium states pertaining to the discrete spectrum, and the other to the continuous one. In the first case, there will be one contribution for each discrete final state, $dR_{n\ell}/dK_\beta$, while the contribution corresponding to a final free Helium state will be dubbed, $dR_{\rm F}/dK_\beta$. The total event rate is given by the sum of all these contributions,
\begin{align}
    \frac{dR}{dK_\beta} = \frac{dR_{\rm F}}{dK_\beta} + \sum_{\ell, n} \frac{dR_{n\ell}}{dK_\beta} \,.
\end{align}
In principle, there are infinitely many discrete states, corresponding to the infinite allowed values for $n$ and $\ell$. In practice, we only computed those contributing to the spectrum within the energy windows shown in Fig.~{\color{Maroon}5} in the main text.
\subsection{Final bound Helium}
\noindent The Helium potential just after decay is predicted to be attractive by both the sudden and semi-sudden approximations. In this case the Helium nucleus can remain bound to the graphene, and the initial and final states of the system can be described by the total wave functions,
\begin{align}
\begin{split}
    \Psi_i ={}& \frac{\chi^{\rm{T}}_{00}(r,z)}{\sqrt{2\pi}} \, , \\
    \Psi_f ={}& \chi^{\mathrm{He}}_{n \ell}(r,z)\frac{e^{i \ell\phi}}{\sqrt{2\pi}} \frac{e^{i \bm{p}_{\beta} \cdot \bm{x}_{\beta}}}{\sqrt{V}} \frac{e^{i \bm{p}_{\nu} \cdot \bm{x}_{\nu}}}{\sqrt{V}} \, ,
\end{split}
\end{align}
with obvious definition of the coordinates.
The matrix element for the transition $\Psi_i \rightarrow \Psi_f$ is then given by,
\begin{align}
\begin{split}
    \mathcal{M}_{f} = \frac{g}{2\pi V} \int_{-\infty}^{+\infty}dz \int_{0}^{+\infty} dr \ r \  \chi_{n \ell}^{\mathrm{He}}(r,z)\chi_{00}^{\mathrm{T}}(r,z) \\
    \times \ e^{-ip_z z} \int_{0}^{2\pi} d\phi \ e^{-i \ell\phi} e^{-ip_{\parallel}r \cos \phi} \, , 
    \label{eq:me1}
\end{split}
\end{align}
where the shorthand notation $p_z = |\bm{p}_{\beta} + \bm{p}_{\nu}| \cos \omega$, $p_{\parallel} = |\bm{p}_{\beta} + \bm{p}_{\nu}| \sin \omega$ has been introduced, $\omega$ being the polar angle of $\bm{p}_{\beta} + \bm{p}_{\nu}$ with respect to the $z$-axis, defined by the normal to the graphene sheet. 

Energy conservation requires that:
\begin{equation}
    m_{\mathrm{T}} + \varepsilon_{00}^{\mathrm{T}} = m_{\mathrm{He}} + \varepsilon_{n \ell}^{\mathrm{He}} + m_e + K_{\beta} + E_{\nu} \, ,
\end{equation}
where $m_{\mathrm{T}}$ and  $m_{\mathrm{He}}$ are nuclear masses, and $\varepsilon_{00}^{\mathrm{T}}$ and  $\varepsilon_{n \ell}^{\mathrm{He}}$ are the energies of the ground state of Tritium and of the excited state of Helium, eigenvalues of the Schr\"{o}dinger equation ({\color{Maroon}14}) of the main text. If we consider a transition to a final state described by the quantum numbers $n, \ell$, the largest electron kinetic energy allowed is,
\begin{align}
    K_{\beta,n \ell}^{\mathrm{end}} \equiv Q - m _{\nu} + \varepsilon_{00}^{\mathrm{T}}-\varepsilon_{n \ell}^{\mathrm{He}} \,,
\end{align}
where $Q \equiv m_{\mathrm{T}}-m_{\mathrm{He}}-m_e$. This corresponds to the region in phase space where the anti-neutrino is emitted with $\bm{p}_{\nu} = 0$. Note that the endpoint of the total $\beta$ spectrum is therefore at $K_{\beta,00}^{\mathrm{end}}$. 

For neutrino mass measurements, one is typically interested in the end-point region of the spectrum. In the present case, we shall  only study the region corresponding to electron kinetic energies $K_{\beta} \in [K_{\beta,00}^{\mathrm{end}}  - 15 \text{ eV}, K_{\beta,00}^{\mathrm{end}}]$. At these energies $|\bm{p}_{\nu}| \ll |\bm{p}_{\beta}|$, so that we may neglect the anti-neutrino momentum appearing in the argument of the exponentials in Eq.~\eqref{eq:me1}. 

We now also exploit the fact that, in this range of energies, the electron momentum is almost constant. Introducing $p_{\beta, nl} \equiv \left[p_{\beta}(K_{\beta,n \ell}^{\mathrm{end}}) + p_{\beta}(K_{\beta,n \ell}^{\mathrm{end}}  - 15 \text{ eV})\right]/2$, we have $p_{\beta} \equiv p_{\beta, n \ell} + \delta p_{\beta}$, where:
\begin{align}
    \delta p _{\beta}(K_{\beta}) = \sqrt{K_{\beta}^2 + 2K_{\beta}m_e} - p_{\beta, n \ell} \, .
\end{align}
Writing $p_{\beta} = p_{\beta, n \ell} ( 1 + \delta p_{\beta}/p_{\beta, n \ell}) \equiv p_{\beta, n \ell} ( 1 + x)$, we now expand in powers of $x(K_{\beta})$, which is $\mathcal{O}(10^{-4})$ for all electron energies considered here. Introducing the auxiliary functions $\alpha(\omega, z) \equiv p_{\beta,n \ell} \ z \cos \omega$ and $\beta(\omega , r) \equiv p_{\beta, n \ell} \ r \sin \omega$, the matrix element then becomes,
\begin{align} \label{eq:mfi}
    \mathcal{M}_{f} \ \equiv \  &\frac{g}{V}(-i)^\ell \Biggl\{ A_{n \ell}(\omega) - xp_{\beta,n \ell}\cos \omega \ B_{n \ell}(\omega) + \notag \\
    & + \frac{1}{2}xp_{\beta,n \ell} \sin \omega \ C_{n \ell}(\omega) \   - i\Bigl[ D_{n \ell}(\omega) \\
    &+ xp_{\beta,n \ell}\cos \omega E_{n \ell}(\omega) + \frac{1}{2}xp_{\beta,n \ell} \sin \omega \ F_{n \ell}(\omega)  \Bigr] \Biggr\} \notag \, ,
\end{align}
where,
\begin{align*}
    A_{n \ell} &\equiv \int dz \int dr \ r \cos \alpha J_\ell(\beta) \chi_{n\ell}^{\mathrm{He}} \chi_{00}^{\mathrm{T}} \, , \\
    B_{n\ell} &\equiv \int dz \int dr \ r \ z \sin \alpha J_\ell(\beta) \chi_{n\ell}^{\mathrm{He}} \chi_{00}^{\mathrm{T}} \, , \\
    C_{n\ell} &\equiv  \int dz \int dr \ r^2 \cos \alpha  [J_{\ell-1}(\beta) - J_{\ell+1}(\beta)]\chi_{n\ell}^{\mathrm{He}} \chi_{00}^{\mathrm{T}} \, , \\
    D_{n\ell} &\equiv \int dz \int dr \ r \sin \alpha J_\ell(\beta) \chi_{n\ell}^{\mathrm{He}} \chi_{00}^{\mathrm{T}} \, , \\
    E_{n\ell} &\equiv \int dz \int dr \ r \ z \cos \alpha J_\ell(\beta)\chi_{n\ell}^{\mathrm{He}} \chi_{00}^{\mathrm{T}} \, , \\
    F_{n\ell} &\equiv \int dz \int dr \ r^2 \sin \alpha [J_{\ell-1}(\beta) - J_{\ell+1}(\beta)]\chi_{n\ell}^{\mathrm{He}} \chi_{00}^{\mathrm{T}} \, ,
\end{align*}
and $J_l(x)$ is the Bessel function of the first kind. Squaring Eq.~\eqref{eq:mfi}, we get:
\begin{equation}
    \bigl| \mathcal{M}_{f} \bigr|^2 = \frac{|g|^2}{V^2} \Bigl[\mathcal{I}_{n\ell}^0(\omega) + x \mathcal{I}_{n\ell}^1(\omega) + \mathcal{O}(x^2) \Bigr] \, ,
\end{equation}
having defined,
\begin{align*}    
    \mathcal{I}_{n\ell}^0 &\equiv A_{n\ell}(\omega)^2 + D_{n\ell}(\omega)^2 \ , \\
    \mathcal{I}_{n\ell}^1 &\equiv 2 \cos \omega \ p_{\beta,n\ell} \bigl[ D_{n\ell}(\omega) E_{n\ell}(\omega) - A_{n\ell}(\omega)B_{n\ell}(\omega)\bigr] \\
    & \ \ \ \ \ + \sin \omega \ p_{\beta, n\ell} \bigl[ A_{n\ell}(\omega)C_{n\ell}(\omega) + D_{n\ell}(\omega)F_{n\ell}(\omega)  \bigr] \, .
\end{align*}
For the $\beta$-decay rate, Fermi's golden rule reads,
\begin{align}
\begin{split}
    d\Gamma_{n\ell} = &2 \pi  \delta \Bigl( E_{\nu} - \bigl[Q-K_{\beta} + \varepsilon_{00}^{\mathrm{T}} - \varepsilon_{n\ell}^{\mathrm{He}} \bigr] \Bigr) \\
    & \times \bigl| \mathcal{M}_{f}(K_{\beta}, \omega) \bigr|^2 \frac{V d^3p_{\beta}}{(2\pi)^3}\frac{V d^3p_{\nu}}{(2\pi)^3} \, ,
\end{split}
\end{align}
so that the differential event rate is,
\begin{align}
\begin{split}
    \frac{dR_{n\ell}}{dK_{\beta}} &= \frac{N_{\mathrm{T}}|g|^2}{4 \pi^3} p_{\nu,n\ell}(K_{\beta})E_{\nu,n\ell}(K_\beta) p_{\beta}E_{\beta} \\
    & \ \ \ \ \ \times \Bigl[ a_{n\ell}^0 + x(K_{\beta})a_{n\ell}^1 + \mathcal{O}(x^2)  \Bigr] \,,
    \label{eq:drbound}
\end{split}
\end{align}
with $N_{\mathrm{T}}$ the number of Tritium atoms and,
\begin{align*}
    a_{n\ell}^0 &\equiv \int_0^\pi d\omega \sin \omega  \ \mathcal{I}_{n\ell}^0(\omega) \,, & a_{n\ell}^1 &\equiv \int_0^\pi d\omega \sin \omega  \ \mathcal{I}_{n\ell}^1(\omega) \, ,
\end{align*}
the zero-th and first order coefficients capturing the features of the transition towards a final state with quantum numbers $n,\ell$. Note, importantly, that these coefficients do not depend on the electron kinetic energy, $K_{\beta}$, so that, once computed, they can be employed in Eq.~\eqref{eq:drbound} to provide an analytical expression for the event rate. Moreover, the power series structure of Eq. (\ref{eq:drbound}) makes manifest the perturbative approach adopted in the computational strategy. As a final remark, notice that $dR_{n,\ell} = dR_{n,-\ell}$. 
\subsection{Final free Helium}
\noindent If the final Helium gains enough energy to be freed from the substrate, we describe the final state of the system as a plane wave,
\begin{align} \label{eq:planewave}
     \Psi_{{f}} ={}& \frac{1}{V^{3/2}} e^{i \bm{p}_{\beta} \cdot \bm{x}_{\beta}} e^{i \bm{p}_{\mathrm{He}} \cdot \bm{x}_{\mathrm{He}}} e^{i \bm{p}_{\nu} \cdot \bm{x}_{\nu}} \, .
\end{align}
In general, the expression above will give the best description of the wave function only for sufficiently large momenta, when the effects of the potential are negligible. For the sake of the present work, Eq.~\eqref{eq:planewave} will suffice.

In addition, to simplify the integration procedure, we  use an approximation for the wave function describing the initial state of the system.  Recall that the initial Tritium is in the ground state of the potential described in Section~{\color{Maroon} 3} of the main text. Near its minimum, this can be approximated with an anisotropic harmonic potential with two fundamental frequencies, one associated to oscillations perpendicular to the graphene layer, and one to oscillations parallel to it. The resulting initial Tritium wave function is an anisotropic Gaussian, 
\begin{align}
    \psi_{i,\rm T}(\bm x) = \frac{1}{\pi^{3/4}\lambda_{\rm p} \sqrt{\lambda_{\rm t}}} \, e^{-\frac{r^2}{2\lambda_{\rm p}^2}-\frac{z^2}{2\lambda_{\rm t}^2}} \,,
\end{align}
where $r$ is the parallel direction, and $z$ the perpendicular one. Correspondingly, $\lambda_{\rm p} \simeq 0.127 \text{ \AA}$ and $\lambda_{\rm t} \simeq 0.084 \text{ \AA}$ are the typical extensions of the wave function in the parallel and transverse directions, respectively. These are extracted from the expansion of the Tritium potential around $\bm x = (0,0,z_0)$. 

The matrix element for the transition is then given by,
\begin{align}
    \mathcal{M}_{f} &   = \frac{g\lambda_p \sqrt{\lambda_t} \, 2^{3/2} \pi^{3/4} }{V^{3/2}} \  e^{-\lambda_t^2 p_z^2/2}   e^{-\lambda_p^2 p_{\parallel}^2/2} \,,
\end{align}
where we have neglected the momentum of the anti-neutrino in the argument of the exponentials. By exploiting conservation of energy, we fix the anti-neutrino energy to be $ E_{\nu} = Q + \varepsilon_{00}^{\mathrm{T}} -  U_{\infty}^\text{He}- K_{\beta} - K_{\mathrm{He}}$. The event rate then reads,
\begin{align*}
\begin{split}
    \frac{dR_{\rm F}}{dK_{\beta}} ={}& \frac{N_{\mathrm{T}}|g|^2\lambda_p^2 \lambda_t}{2 \pi^{7/2}} p_{\beta} E_{\beta} \int_0^{p_{\mathrm{He}}^{\mathrm{max}}} dp_{\mathrm{He}} \  p_{\mathrm{He}}^2 \int_0^{\pi} d \omega \sin \omega  \\
    & \times \int_0^{\pi} d \alpha \sin \alpha\  p_{\nu}E_{\nu}\ I_0 \bigl( 2\lambda_p^2p_{\beta} p_{\mathrm{He}} \sin \omega \sin \alpha \bigr) \\
    & \times e^{-\lambda_t^2(p_{\beta} \cos \omega + p_{\mathrm{He}} \cos \alpha)^2} e^{-\lambda_p^2(p_{\beta}^2 \sin^2 \omega + p_{\mathrm{He}}^2 \sin^2 \alpha)} \, ,
\end{split} 
\end{align*}
where $I_0$ is the modified Bessel function of the first kind, and $p_{\mathrm{He}}^{\mathrm{max}} \equiv \sqrt{2m_{\mathrm{He}}(Q+\varepsilon_{00}^{\mathrm{T}}- U_{\infty}^\text{He}-K_{\beta}-m_{\nu})}$ is the maximum Helium momentum obtained, by requiring that $E_{\nu} \geq m_\nu$.
\bibliographystyle{apsrev4-1}
\bibliography{biblio.bib}